# Thermal Fluctuations of Matter Composition and Quark Nucleation in Compact Stars

Mirco Guerrini[1,2], Giuseppe Pagliara[1,2], Alessandro Drago[1,2], and Andrea Lavagno[3,4]
[1] Department of Physics and Earth Science, University of Ferrara, 44122 Ferrara, Italy
[2] INFN, Sezione di Ferrara, 44122 Ferrara, Italy
[3] Department of Applied Science and Technology, Politecnico di Torino, 10129 Torino, Italy
[4] INFN Sezione di Torino, 10125 Torino, Italy

## Abstract

At the extreme densities reached in the core of neutron stars, it is possible that deconfined quark matter is produced. The formation of this new phase of strongly interacting matter is likely to occur via a first-order phase transition for the typical temperatures reached in astrophysical processes. The first seeds of quark matter would then form through a process of nucleation within the metastable hadronic phase. Here, we address the role of the thermal fluctuations in the hadronic composition on the nucleation of two-flavor quark matter. At finite temperature, the thermodynamic quantities in a system fluctuate around average values. Nucleation being a local process, it is possible that it occurs in a subsystem whose composition makes the nucleation easier. We will consider the total probability of the nucleation as the product between the probability that a subsystem has a certain hadronic composition different from the average in the bulk, and the nucleation probability in that subsystem. We will show how those fluctuations of the hadronic composition can increase the efficiency of nucleation already for temperatures $\sim(0.1-1)$ keV. However, for temperatures $\lesssim(1-10)$ MeV, the needed overpressure exceeds the maximum pressure reached in compact stars. Finally, for even larger temperatures the process of nucleation can take place, even taking into account finite-size effects.

*Unified Astronomy Thesaurus concepts:* Neutron star cores (1107); Neutron stars (1108); Compact objects (288); Nuclear physics (2077); Nuclear astrophysics (1129)

## 1. Introduction

Quantum chromodynamics (QCD), namely the theory that describes strongly interacting matter, predicts that at sufficiently high baryonic densities, hadronic matter undergoes a phase transition to deconfined quark matter. The order of this phase transition and its critical density are, however, totally unknown. Neutron stars (NSs), being the densest stellar objects in the Universe, are the most promising sites for this transition to occur, and in recent years several studies have shown that by using the presently available observational data, the core of NSs is very likely to be formed by deconfined quark matter (see, e.g., Annala et al. 2020). The more extreme scenario in which stars entirely formed by quark matter exist, namely quark stars, is also viable and actually favored by recent data indicating that the maximum mass of compact stars could be larger than $\sim 2.6\,M_\odot$ (Bombaci et al. 2021).

A crucial question concerns the astrophysical evolutionary paths that lead to the formation of deconfined quark matter. A possible scenario has been proposed in Fischer et al. (2018): Deconfinement already occurs during the initial stages of the core-collapse supernovae (CCSNe) associated with blue supergiant stars. Actually, it is the formation of quark matter itself that provides the necessary energy output that leads to the explosion; see Sagert et al. (2009). Another possibility is that deconfinement occurs after a proto-neutron star (PNS) has been formed, and specifically only when neutrino untrapping sets in (Pons et al. 2001). Finally, binary neutron star mergers (BNSMs) could also possibly produce quark matter inside the hot and fast-rotating post-merger remnant (Bauswein et al. 2019; Prakash et al. 2021). In all these studies, quark matter is assumed to be produced in thermodynamic equilibrium with the hadronic phase, and the microphysics of the production of finite-size quark matter structures is neglected.

However, if deconfinement is a first-order phase transition, the process of nucleation of quark matter droplets within the metastable hadronic phase must be taken into account in order to provide a complete description of the system. By considering that in CCSNe, PNSs, and BNMs temperatures up to a few tens of MeV can be reached, thermal nucleation is most likely the more efficient mechanism at work for the formation of quark matter droplets. Nucleation of quark matter in NSs has been studied in a number of papers and within two approaches: (i) nucleation of $\beta$-stable quark matter within $\beta$-stable hadronic matter (see, e.g., Berezhiani et al. 2003; Drago et al. 2004; Mintz et al. 2010a, 2010b), and (ii) nucleation of the quark phase out of chemical equilibrium within $\beta$-stable hadronic matter (see, e.g., Iida & Sato 1997; Olesen & Madsen 1994; Iida & Sato 1998; Bombaci et al. 2004, 2009, 2016). The second approach, in particular, is based on the following argument: The nucleation of quark matter is a process mediated by the strong interaction, whose typical timescale ($\sim 10^{-23}$ s) is much smaller than that of the weak interaction. Thus, during the formation of the first seeds of quark matter, weak interactions do not have sufficient time to change the flavor composition of matter. The nucleation process, therefore, is calculated by using an out-of-equilibrium quark phase whose flavor composition is identical to that of the $\beta$-stable hadronic phase and not by taking into account the quark phase already in $\beta$-equilibrium. Only after the creation of the first seed of quark matter do weak interactions have time to change the composition of quark matter and bring it to $\beta$-equilibrium.

NS matter is a multicomponent system in which thermal fluctuations in the number densities of the different particles could be significant, at least at high temperatures. In the first of the two approaches outlined above, it is assumed that those fluctuations are so efficient that quark matter is produced already at chemical equilibrium, while in the second approach the role of those fluctuations is completely disregarded. The aim of this work is to address quantitatively the role of the thermal fluctuations of number densities on the nucleation of quark matter. To this purpose, we will develop a general framework for dealing with the nucleation associated with first-order phase transitions in multicomponent systems. In this paper, we will limit the discussion to the case of two-flavor quark and hadronic matter (i.e., nucleonic matter), thus to a two-component system, and we leave the study of the three-flavor case for a forthcoming paper.[5]

The paper is organized as follows. In Section 2, a brief theoretical background for calculating fluctuations and nucleation is presented. The description of the nucleation process in the presence of thermal fluctuations is detailed in Section 3. The equations of state (EOSs) for nucleons, quarks and electrons are given in Section 4. Results of our calculations are presented in Section 5. A summary and the conclusions are presented in Section 6.

## 2. Background

In this section, we present a brief theoretical background on thermal fluctuations in the hadronic phase and on nucleation.

### 2.1. Thermal Fluctuations

Let us consider a closed macroscopic system. Let us focus on a certain subsystem, small compared to the whole system, in which the thermodynamic quantities fluctuate. According to Boltzmann's principle, the probability of a fluctuation of one or more thermodynamical quantities is

$$\mathcal{P} = K \exp[-W_{\min}/T_0], \quad (1)$$

where $W_{\min}$ is the minimum work needed to produce a certain variation of the thermodynamic quantities in the considered subsystem (namely the work needed if the transformation is reversible) and $K$ the normalization factor (Landau & Lifshitz 1969). The work done is equal to the variation of the total internal energy of the whole system:

$$W_{\min} = \Delta E + \Delta E_0, \quad (2)$$

where $E$ is the internal energy of the small subsystem and $E_0$ is the internal energy of the surrounding part, which plays the role of a reservoir (of energy and/or particles). Since the reservoir is much larger than the subsystem, we can assume that the intensive thermodynamic quantities such as the pressure $P_0$, the temperature $T_0$, and the chemical potentials $\{\mu_{i,0}\}$ of all the particle species $i$ remain constants. Note that no assumptions at this stage are made on the thermodynamic quantities in the subsystem. These considerations allow us to substitute the following relation:

$$\Delta E_0 = T_0 \Delta S_0 - P_0 \Delta V_0 + \sum_i \mu_{i,0} \Delta N_{i,0}. \quad (3)$$

in the minimum work formula, leading to

$$W_{\min} = \Delta E + T_0 \Delta S_0 - P_0 \Delta V_0 + \sum_i \mu_{i,0} \Delta N_{i,0}. \quad (4)$$

Using the reversibility of the process $\Delta S = -\Delta S_0$, the conservation of the total volume $\Delta V = -\Delta V_0$, and the conservation of the number of particles for each species $\Delta N_i = -\Delta N_{i,0}$, we obtain (Landau & Lifshitz 1969)

$$W_{\min} = \Delta E - T_0 \Delta S + P_0 \Delta V - \sum_i \mu_{i,0} \Delta N_i. \quad (5)$$

Note that the conservation of the number of particles for each species used in the previous step is not necessarily true. For example, in the case in which weak interactions play a role, the number of particles in the system is not conserved for all the species $i$ independently. However, we can in principle rephrase the above discussion by considering $N_i$ as being the (net) numbers of the globally conserved charges (e.g., if weak interactions occur the baryon number $B$ and the electric charge $C$ are conserved while the strangeness $S$ and the isospin $I$ are not).

### 2.2. Nucleation

Nucleation can be considered as a particular case of fluctuation, and thus it can be studied within the same formalism introduced before. In a homogeneous phase, small localized fluctuations in the thermodynamical variables can give rise to the appearance of virtual drops of a new phase (e.g., a liquid droplet in a vapor phase or a quark droplet in a hadronic phase). If the homogeneous phase is stable, these droplets are unstable and disappear. However, if the homogeneous phase is metastable (i.e., if it is not the most energetically convenient bulk phase) when the fluctuation generates a droplet large enough to be stable, this first seed could trigger the transformation of all or part of the homogeneous system into the new phase (Landau & Lifshitz 1969). In order to trigger the phase transition, a droplet of the new phase in the metastable homogeneous phase has to be "large enough," since the gain in terms of volume energy (i.e., in terms of bulk energy) has to be able to overcome the energy needed to create the interface between the two phases. Assuming that the seed is spherical, the critical radius $R_c$ is the minimal radius for a droplet of the new phase to be in unstable equilibrium with the homogeneous phase, leading then to the phase transition. The conditions for the existence of a seed of the new phase "II" in (unstable) equilibrium with the old homogeneous phase "I" are the following (Landau & Lifshitz 1969):

$$P_{\mathrm{I}} = P_{\mathrm{II}} - \frac{2\sigma}{R_c}, \quad (6)$$

$$\mu_{\mathrm{I},k}(P_{\mathrm{I}}, T_{\mathrm{I}}) = \mu_{\mathrm{II},k}(P_{\mathrm{II}}, T_{\mathrm{II}}), \quad (7)$$

$$T_{\mathrm{I}} = T_{\mathrm{II}}, \quad (8)$$

namely the mechanical, chemical, and thermal equilibrium conditions, where $\sigma$ is the surface tension and $\mu_k$ the conserved chemical potentials, one for every globally conserved charge.

---
[5] The two conserved charges in the system are the baryonic and the electric charges, which can be mapped into the conservation of up and down quark flavors. Electrons are also included in the calculations but their density is fixed by the requirement of charge neutrality.

The presence of a finite-size term in the mechanical equilibrium condition implies that $P_{II} > P_I$. This also implies that the transition will start at a pressure $P_I > P_x$, where $P_x$ is the equilibrium pressure when $\sigma \to 0$. We will call overpressure the difference $P_I - P_x$.

## 3. Framework

At finite temperature, the thermodynamic quantities in a system fluctuate around average values. In particular, let us consider a hadronic system at fixed temperature and pressure. It is possible to compute the average values of the particle fractions in $\beta$-equilibrium $\{y_i^{H_\beta}\}$.[6] The values obtained, however, are just average values. We will call $H_\beta$ a hadronic system in $\beta$-equilibrium. If we divide the system into several small subsystems, we cannot assume that the actual composition $\{y_i^{H^*}\}$ in each of these subsystems is identical to the average one. We will call $H^*$ an out-of-equilibrium hadronic subsystem having the composition $\{y_i^{H^*}\}$.

We expect that the higher the temperature, the more likely it is to find subsystems with compositions that significantly deviate from the averages. Since nucleation is a local process, it is possible that it occurs in a subsystem whose composition makes the formation of a seed of the new phase more convenient than if average values are taken into account.

In this scheme, the probability of generating a seed of the new phase is the product of two probabilities, $\mathcal{P}_1 \mathcal{P}_2$, where $\mathcal{P}_1$ is the probability that a certain subsystem of the hadronic phase is in the $H^*$ state, whose particle fractions $\{y_i^{H^*}\}$ are different from the average fractions $\{y_i^{H_\beta}\}$ by $\{\Delta y_i\}$:

$$y_i^{H^*} = y_i^{H_\beta} + \Delta y_i, \quad (9)$$

and $\mathcal{P}_2$ is the probability that a critical droplet of $Q^*$ quark matter is nucleated from a subsystem in which the hadronic phase is in a $H^*$ state. The flavor composition $\{y_i^{Q^*}\}$ of $Q^*$ is "frozen," namely it corresponds to $\{y_i^{H^*}\}$ for the same argument as to the strong and weak interaction timescales reported in Section 1:

$$y_i^{Q^*} = y_i^{H^*}. \quad (10)$$

Thus, we define $Q^*$ as an out-of-equilibrium quark phase having the same flavor composition of $H^*$.

### 3.1. Nucleation ($H^* \to Q^*$)

Let us start by computing the probability $\mathcal{P}_2$. The first key element is the work $W_2$, namely the minimum work needed to generate a seed of quark matter $Q^*$ from a subsystem $H^*$ of the hadronic system. As discussed above, and following Bombaci et al. (2016), the flavors are conserved during nucleation. For example, in the two-flavor case,

$$N_u^{Q^*} = N_u^{H^*} = 2N_p^{H^*} + N_n^{H^*}, \quad (11)$$

$$N_d^{Q^*} = N_d^{H^*} = N_p^{H^*} + 2N_n^{H^*}, \quad (12)$$

$$N_e^{Q^*} = N_e^{H^*}. \quad (13)$$

This implies also that the total baryonic number is conserved $N_p^{H^*} + N_n^{H^*} = N_B^{H^*} = N_B^{Q^*} = (N_u^{Q^*} + N_d^{Q^*})/3 = N_B$, thus

$$y_u^{Q^*} = y_u^{H^*} = 2y_p^{H^*} + y_n^{H^*}, \quad (14)$$

$$y_d^{Q^*} = y_d^{H^*} = y_p^{H^*} + 2y_n^{H^*}, \quad (15)$$

$$y_e^{Q^*} = y_e^{H^*}, \quad (16)$$

where $y_i = N_i/N_B$ are the particle fractions that we will use in the following calculations. Note that $H^*$ is in principle an out-of-equilibrium phase, thus $\mu_e \neq -\mu_C$. By using Equation (5) and by introducing the free energy $F = E - ST$ and the Gibbs free energy $\Phi = F + PV$, we obtain

$$\begin{aligned} W_2 &= F_{Q^*} - F_{H^*} + P_{H_\beta}(V_{Q^*} - V_{H^*}) + \sigma \mathcal{S}_{Q^*} \\ &= \Phi_{Q^*} - \Phi_{H^*} - V_{Q^*}(P_{Q^*} - P_{H_\beta}) \\ &\quad + V_{H^*}(P_{H^*} - P_{H_\beta}) + \sigma \mathcal{S}_{Q^*} \\ &= \Phi_{Q^*} - \Phi_{H^*} - V_{Q^*}(P_{Q^*} - P_{H^*}) + \sigma \mathcal{S}_{Q^*}, \end{aligned} \quad (17)$$

where $\sigma$ is the surface tension and $\mathcal{S}_{Q^*}$ the surface area of the quark matter droplet (see Landau & Lifshitz 1969). We have assumed the temperature to be constant in the whole system during the process $T_{H_\beta} = T_{H^*} = T_{Q^*}$ and that the surrounding part "0" is in the hadronic phase. Moreover, $P_{H^*} = P_{H_\beta} = P_H$ since we are assuming that the hadronic matter fluctuation occurs at constant pressure (see Section 3.2). The surface tension between the hadronic and quark phases $\sigma$ plays a key role in the phase transition by increasing/decreasing the work required to form a quark matter droplet. If too large, it can even prevent quark nucleation (see, e.g., Mintz et al. 2010a, 2010b). However, its value is poorly known at present and estimates for its value are strongly model dependent. In Heiselberg et al. (1993), it was found that $\sigma \sim (10-100)$ MeV fm$^{-2}$. In Iida & Sato (1997, 1998), a value of $\sigma \sim 10$ MeV fm$^{-2}$ was estimated by using a Fermi-gas model for a quark matter droplet in vacuum. In Fraga et al. (2019), lower values were obtained, $\sigma \sim (2-10)$ MeV fm$^{-2}$, by using a chiral nucleon-meson model (without quark degrees of freedom) in which the chirally restored phase is identified with quark matter. Moreover, $\sigma$ could also depend on the temperature. For instance, in Bessa et al. (2009), the temperature dependence of $\sigma$ is consistently computed within a general framework in which a potential for the order parameter is introduced to describe first-order phase transitions. In the present work, we will assume $\sigma$ to be a density- and temperature-independent input parameter. Its impact will be investigated by using two different values $\sigma = 10, 30$ MeV fm$^{-2}$.

The "local" compositions $\{y_i^{H^*}\}$ and $\{y_i^{Q^*}\}$ are related to the average compositions by Equations (9)–(10) and, under the assumption that the temperature $T$ is the same in all the phases, we are left with the following independent variables: the composition fluctuations $\{\Delta y_i\}$, the temperature $T$, the quark seed volume and surface $V_{Q^*}$ and $\mathcal{S}_{Q^*}$, and two more intensive variables, one for each phase (e.g., $P_{Q^*}$ and $P_{H^*}$ or $n_{B,Q^*}$ and $n_{B,H^*}$, …). However, we want to rewrite the system only in terms of hadronic quantities as independent variables. The idea is then to use the equilibrium conditions to connect the intensive independent quark variable with the hadronic one. We will follow the approach of Bombaci et al. (2016; see also Landau & Lifshitz 1969), where a low degree of metastability $P_{H^*} \approx P_{Q^*}$ is assumed. We will say that the system has a "low degree of metastability" if

$$\delta P_{H^*} = |P_{H*} - P_x| \ll P_x, \quad (18)$$

$$\delta P_{Q^*} = |P_{Q^*} - P_x| \ll P_x, \quad (19)$$

where $P_x$ is, again, the pressure at the equilibrium when $R_c \to +\infty$ (plane surface) or $\sigma \to 0$, so that

$$\mu_{H^*,k}(P_x, T) = \mu_{Q^*,k}(P_x, T), \quad (20)$$

where $k$ labels every globally conserved charge. That is, $P_x$ is the common pressure of the two phases in an ordinary first-order phase transition at temperature $T$. In other words, a system has a low degree of metastability if the overpressure needed in the metastable phase to balance the finite-size effects due to the surface tension is relatively small.

---

[6] Here, $i$ labels all the particle species in the hadronic phase (i.e., $i = p, n, e$ in the two-flavor case) or, equally, a set of a linear combination of them. For example, in the two-flavor case, one can use $y_p, y_n$ as independent variables or $y_u = 2y_p + y_n$, $y_d = y_p + 2y_n$, namely the quark flavor composition. The fraction of a particle species $i$ is $y_i = N_i/N_B$ where $N_i$ and $N_B$ are the numbers of particles $i$ and the baryon number. If not otherwise specified, $N_i$ and $y_i$ will stay, respectively, for the net number and fraction of $i$ (e.g., $y_e = y_{e^-} - y_{e^+}$) while the other thermodynamic quantities for the sum of the two contributions (e.g., $P_e = P_{e^-} + P_{e^+}$).

Under the condition of low metastability, the quark Gibbs energy $\Phi_{Q^*}(P_{Q^*}, T)$ can be expanded around $P_{H^*}$:

$$\Phi_{Q^*}(P_{Q^*}, T) \simeq \Phi_{Q^*}(P_{H^*}, T) + \left.\frac{\partial \Phi_{Q^*}}{\partial P_{Q^*}}\right|_T (P_{Q^*} - P_{H^*})$$

$$\simeq \Phi_{Q^*}(P_{H^*}, T) + V_{Q^*}(P_{Q^*} - P_{H^*}). \quad (21)$$

By substituting Equation (21) in Equation (17), we obtain

$$W_2(P_{H^*}, T) = \Phi_{Q^*}(P_{H^*}, T) - \Phi_{H^*}(P_{H^*}, T) + \sigma \mathcal{S}_{Q^*}. \quad (22)$$

Note that $\Phi_{Q^*}$ is computed at the hadronic pressure $P_{H^*}$. Using $\Phi = \sum_i \mu_i N_i$, we have

$$W_2 = \sum_i \mu_i^{Q^*} N_i^{Q^*} - \sum_i \mu_i^{H^*} N_i^{H^*} + \sigma \mathcal{S}_{Q^*}$$

$$= N_B \left[ \sum_i \mu_i^{Q^*} y_i^{Q^*} - \sum_i \mu_i^{H^*} y_i^{H^*} \right] + \sigma \mathcal{S}_{Q^*}$$

$$= n_{B,Q^*} V_{Q^*}(\mu_{Q^*} - \mu_{H^*}) + \sigma \mathcal{S}_{Q^*}, \quad (23)$$

where

$$\mu_{H^*} = \frac{\Phi_{H^*}}{N_B} = \frac{P_{H^*} + \varepsilon_{H^*} - s_{H^*}T}{n_{B,H^*}} = \sum_i \mu_i^{H^*} y_i^{H^*} \quad (24)$$

is the average chemical potential (or Gibbs energy per baryon) of the hadronic phase, and a similar expression holds for the $Q^*$ phase. It is important to note that in Equation (23) the thermodynamic quantities are computed at the same (external) pressure. Another possible approach is presented in Mintz et al. (2010a, 2010b), where the low degree of metastability is not assumed, but the hadron and quark phases are computed at the same Gibbs energy per baryon. In that case, we would obtain $W_2 = -V_{Q^*}(P_{Q^*} - P_{H^*}) + \sigma \mathcal{S}_{Q^*}$. Finally, note that Equation (23) can also be obtained starting from a free energy functional of the order parameter with an effective potential having a global minimum and a local minimum (metastable phase; Langer 1969; Scavenius et al. 2001; Bessa et al. 2009; Mintz et al. 2010b). For example, in Bessa et al. (2009), in the spirit of the Landau–Ginzburg theory, an effective potential is introduced for the order parameter (modeled as a scalar field). By solving the corresponding Euler–Lagrange equation, under the assumption of a rapid variation of the field within the region separating the metastable phase from the stable phase, Equation (23) is obtained. That corresponds to the so-called thin-wall approximation, which we also assume here, that allows one to neglect the width of the surface separating the two phases with respect to the radius of the quark matter droplet. Moreover, in Bessa et al. (2009), the thin-wall approximation is applied using both a constant and a temperature-dependent surface tension, $\sigma(T)$. In particular, it has been shown that if the temperature dependence of $\sigma$ is neglected, the nucleation rates are overestimated. This could indeed affect our results, which are instead obtained with constant $\sigma$. This issue needs to be investigated in future works.

Assuming that the quark seed is a sphere of radius $R$, we obtain

$$W_2 = \frac{4}{3}\pi R^3 n_{B,Q^*}(\mu_{Q^*} - \mu_{H^*}) + 4\pi\sigma R^2. \quad (25)$$

The free variables are then $P_H$, $\{\Delta y_i\}$, $T$, and $R$. As reported in Section 4, the used EOSs are in the form

$$X = X(n_B, \{y_i\}, T), \quad (26)$$

where $X = \varepsilon, P, \mu, \ldots$ is a generic thermodynamic quantity. The Gibbs energies per baryon of $Q^*$ and $H^*$ are then computed as

$$\mu_{Q^*}(P_H, \{\Delta y_i\}, T) = \mu_Q[n_{B,Q}(P_H, \{y_i^{Q^*}\}, T), \{y_i^{Q^*}\}, T], \quad (27)$$

$$\mu_{H^*}(P_H, \{\Delta y_i\}, T) = \mu_H[n_{B,H}(P_H, \{y_i^{H^*}\}, T), \{y_i^{H^*}\}, T], \quad (28)$$

where $n_{B,Q^*} = n_{B,Q}(P_H, \{y_i^{Q^*}\}, T)$ and $n_{B,H^*} = n_{B,H}(P_H, \{y_i^{H^*}\}, T)$ are the baryon densities in the quark and hadron phases at which the pressure is $P_H$. The free variables $\{\Delta y_i\}$ are implicitly contained in $\{y_i^{Q^*}\}$ and $\{y_i^{H^*}\}$. The particle fractions are computed at the same pressure:

$$y_u^{Q^*}(P_H, \{\Delta y_i\}, T) = 2y_p^{H^*}(P_H, \{\Delta y_i\}, T) + y_n^{H^*}(P_H, \{\Delta y_i\}, T), \quad (29)$$

$$y_d^{Q^*}(P_H, \{\Delta y_i\}, T) = y_p^{H^*}(P_H, \{\Delta y_i\}, T) + 2y_n^{H^*}(P_H, \{\Delta y_i\}, T), \quad (30)$$

$$y_e^{Q^*}(P_H, \{\Delta y_i\}, T) = y_e^{H^*}(P_H, \{\Delta y_i\}, T). \quad (31)$$

The work has two terms: a volume term $\propto R^3$ that can be both positive and negative, and a surface term $\propto R^2$ that is always positive and is due to the finite-size effects. The work $W_2$ is then positive and growing for all $R$ if $\mu_{Q^*}(P_H, T) > \mu_{H^*}(P_H, T)$, while it reaches a maximum and then decreases if $\mu_{Q^*}(P_H, T) < \mu_{H^*}(P_H, T)$, i.e., if the hadronic phase is metastable. In such a case, $W_2$ is a potential barrier. The critical radius is the radius at which $W_2$ has a maximum:

$$R_c(P_H, \{\Delta y_i\}, T) = \frac{2\sigma \, n_{B,Q^*}^{-1}(P_H, \{\Delta y_i\}, T)}{\mu_{H^*}(P_H, \{\Delta y_i\}, T) - \mu_{Q^*}(P_H, \{\Delta y_i\}, T)}. \quad (32)$$

The work at the critical radius is

$$W_{2,c}(P_H, \{\Delta y_i\}, T) = \frac{16\pi}{3} \frac{\sigma^3 \, n_{B,Q^*}^{-2}(P_H, \{\Delta y_i\}, T)}{[\mu_{H^*}(P_H, \{\Delta y_i\}, T) - \mu_{Q^*}(P_H, \{\Delta y_i\}, T)]^2}. \quad (33)$$

We want now to compute the probability of overcoming the potential barrier and generate the first seed of quark matter. There are two possible mechanisms: thermal fluctuations or quantum tunneling.

The thermal nucleation probability (Langer 1969; Langer & Turski 1973; Landau & Lifshitz 1969), namely the probability of generating a critical seed of quark matter via thermal nucleation, reads

$$\mathcal{P}_2^{\mathrm{th}}(P_H, \{\Delta y_i\}, T) = \exp\left[-\frac{W_{2,c}(P_H, \{\Delta y_i\}, T)}{T}\right]. \quad (34)$$

The quantum tunneling nucleation will be treated within a semiclassical approach (Iida & Sato 1998). First, we will compute the ground-state energy $E_0$ of the drop in the potential barrier $W_2$ in the Wentzel–Kramers–Brillouin (WKB) approximation. Then, the probability of tunneling will be given by

$$\mathcal{P}_2^{\mathrm{qt}}(P_H, \{\Delta y_i\}, T) = \exp[-A_2(P_H, \{\Delta y_i\}, T)], \quad (35)$$

where $A_2(P_H, \{\Delta y_i\}, T)$ is the action under the barrier computed at the ground-state energy $E_0(P_H, \{\Delta y_i\}, T)$:

$$A_2(E) = 2\int_{R_-}^{R_+} \sqrt{[2\mathcal{M}(R) + E - W_2(R)][W_2(R) - E]} \, dR, \quad (36)$$

where $R_-$ and $R_+$ are the classical turning points and $\mathcal{M}$ the droplet effective mass. All details can be found in Iida & Sato (1998).

### 3.2. Composition Fluctuation ($H_\beta \to H^*$)

Finally, let us compute the probability $\mathcal{P}_1$. Following Landau & Lifshitz (1969), we will compute the probability that at constant pressure and temperature and for a fixed number $N_B$ of baryons in the subsystem, the local particle composition of hadronic matter differs from its average values by $\{\Delta y_i\}$. As reported in Section 2.1, this probability is

$$\mathcal{P}_1 = K \exp\left[-\frac{W_1}{T}\right], \quad (37)$$

where $K$ is the normalization factor. The work $W_1$ is the minimum work needed to change the hadronic composition by a set $\{\Delta y_i\}$:

$$W_1 = F_{H^*} - F_{H_\beta} + P_{H_\beta}(V_{H^*} - V_{H_\beta}) - N_B \sum_i \mu_i^{H_\beta} \Delta y_i, \quad (38)$$

where we have used Equation (5) and $\Delta y_i = y_i^{H^*} - y_i^{H_\beta}$. Then,

$$W_1 = \Phi_{H^*} - \Phi_{H_\beta} - N_B \sum_i \mu_i^{H_\beta} \Delta y_i \quad (39)$$

$$= N_B \sum_i y_i^{H^*}(\mu_i^{H^*} - \mu_i^{H_\beta}), \quad (40)$$

where we used $P_{H_\beta} = P_{H^*} = P_H$ since we are computing the fluctuations at a given constant pressure and $\Phi = F + PV = \sum_i \mu_i N_i$. The chemical potentials $\mu_i$ in $H^*$ and in $H_\beta$ are computed at the same fixed pressure and temperature:

$$\mu_i^{H^*}(P_H, \{\Delta y_i\}, T) = \mu_i^H[n_{B,H}(P_H, \{y_i^{H^*}\}, T), \{y_i^{H^*}\}, T]$$
$$\mu_i^{H_\beta}(P_H, T) = \mu_i^H[n_{B,H}(P_H, \{y_i^{H_\beta}\}, T), \{y_i^{H_\beta}\}, T],$$

where $n_{B,H^*} = n_{B,H}(P_H, \{y_i^{H^*}\}, T)$ and $n_{B,H_\beta} = n_{B,H}(P_H, \{y_i^{H_\beta}\}, T)$ are the baryon densities computed at the same pressure and temperature but at different particle fractions (i.e., different compositions). Again, $y_i^{H^*} = y_i^{H_\beta} + \Delta y_i$. The independent variables at this stage are then $N_B$, $P_H$, $T$, and $\{\Delta y_i\}$. The quantity $W_1(N_B, \{\Delta y_i\}, P_H, T)$ is always positive, and it is zero when $\Delta y_i = 0$ (i.e., when the composition of the considered subsystem is equal to the average composition of the bulk, thus $\mu_i^{H^*} = \mu_i^{H_\beta}$).

The greater the number of particles fluctuating $\{\Delta N_i\} = \{N_B \Delta y_i\}$, the greater the work needed for such a fluctuation $W_1$ and the lower the probability $\exp(-W_1/T)$ for it to occur. For vanishingly small temperature, $\exp(-W_1/T) \neq 0$ only if $\{\Delta y_i\} \to 0$, thus the role of thermal fluctuation in the hadronic composition is negligible. Moreover, it can be shown that in the small fluctuation limit ($\{\Delta y_i\} \ll 1$, $\{\Delta N_i\} \ll N_B$), $\exp(-W_1/T)$ is a multivariate Gaussian (see Appendix A).

The number of baryons $N_B$ we are interested in is the one that is contained in the seed of quark matter. Note that $N_{B,H_\beta} = N_{B,H^*} = N_{B,Q^*} = N_B$ by construction, but $V_{H_\beta} \neq V_{H^*} \neq V_{Q^*} = 4/3\pi R^3$. We fix $N_B$ by using the volume of the first quark droplet:

$$N_B(R, P_H, \{\Delta y_i\}, T) = n_{B,Q^*}(P_H, \{\Delta y_i\}, T) V_{Q^*}(R). \quad (41)$$

Thus,

$$W_1 = n_{B,Q^*} V_{Q^*} \sum_i y_i^{H^*}(\mu_i^{H_\beta} - \mu_i^{H^*}) \quad (42)$$

$$= n_{B,Q^*} \frac{4}{3}\pi R^3 \sum_i y_i^{H^*}(\mu_i^{H_\beta} - \mu_i^{H^*}). \quad (43)$$

In the thermal nucleation case, the first droplet of quark matter has a critical radius $R_c$, thus

$$W_{1,c}(P_H, \{\Delta y_i\}, T) = W_1[R_c(P_H, \{\Delta y_i\}, T), P_H, \{\Delta y_i\}, T], \quad (44)$$

where the critical radius $R_c$ is computed in Equation (32). Whereas, in the quantum nucleation case, the first droplet of quark matter is generated with a radius $R_+$, namely the classical turning point, thus

$$W_{1,+}(P_H, \{\Delta y_i\}, T) = W_1[R_+(P_H, \{\Delta y_i\}, T), P_H, \{\Delta y_i\}, T]. \quad (45)$$

The classical turning point is computed as shown in Iida & Sato (1998).

We finally need to identify the normalization factor. In principle, $K(P_H, T)$ should be computed by imposing the multidimensional integral of $\mathcal{P}_1(P_H, \{\Delta y_i\}, T)$ (Equation (37)) in $-\infty < \{\Delta y_i\} < +\infty$ (i.e., a multidimensional integral where the integrating variables are all $\Delta y_i$) be equal to 1. An estimation of the normalization factor will be provided in Appendix A. However, as will be shown in Section 5, the nucleation probability varies exponentially as pressure and temperature change, while the normalization factor $K(P_H, T)$ has a much weaker dependence on them. The role of the normalization factor is then negligible in order to identify the temperature and pressure conditions at which nucleation occurs. Thus, we will set $K(P_H, T) = 1$ for simplicity:

$$\mathcal{P}_1^{\text{th}}(P_H, \{\Delta y_i\}, T) = \exp\left[-\frac{W_{1,c}(P_H, \{\Delta y_i\}, T)}{T}\right], \quad (46)$$

$$\mathcal{P}_1^{\text{qt}}(P_H, \{\Delta y_i\}, T) = \exp\left[-\frac{W_{1,+}(P_H, \{\Delta y_i\}, T)}{T}\right]. \quad (47)$$

### 3.3. Total Process ($H_\beta \rightarrow H^* \rightarrow Q^*$)

The total probability to generate a seed of $Q^*$ matter in a hadronic matter with a certain pressure $P_H$ and temperature $T$ is thus given by the product between the (quantum or thermal) nucleation probability $\mathcal{P}_2$ and the probability $\mathcal{P}_1$ that the considered hadronic subsystem, characterized by a number of baryons $N_B$, has a composition $\{y_i^{H^*}\}$. The number of baryons $N_B$ in Equation (40) is fixed by the critical radius $R_c$ or by the classical turning point radius $R_+$ in the thermal and quantum nucleation, respectively, as shown in Equations (44) and (45).

A useful quantity to compute is the nucleation time, namely the typical time needed to generate a critical seed of quark matter. In the case of thermal nucleation (Langer 1969; Langer & Turski 1973), it reads

$$\tau^{\text{th}}(P_H, \{\Delta y_i\}, T) = \left[V_{\text{nuc}} \frac{\kappa}{2\pi} \Omega_0 \mathcal{P}_1 \text{th} \mathcal{P}_2 \text{th}\right]^{-1}, \quad (48)$$

where $\kappa$ is the so-called dynamical prefactor, which is related to the growth rate of the drop radius $R$ near the critical radius $R_c$, and $\Omega_0$ is the so-called statistical prefactor, which measures the phase-space volume of the saddle-point region around $R_c$. The result of multiplying $\mathcal{P}_2$ by the prefactors is the thermal nucleation rate, namely the number of critical quark seeds generated in one cubic femtometer per second. Here, $V_{\text{nuc}}$ is the volume inside which the values of $P_H$ and $T$ are approximately equal to the ones used in the calculation and where the nucleation has the highest probability of taking place. Usually, for compact objects (Bombaci et al. 2016), this region is assumed to be a sphere of ∼100 m in the center of the star.

If, instead, quantum nucleation is the fastest process (Iida & Sato 1998), the nucleation time reads

$$\tau^{\text{qt}}(P_H, \{\Delta y_i\}, T) = [N_{\text{nuc}} \nu_0 \mathcal{P}_1^{\text{qt}} \mathcal{P}_2^{\text{qt}}]^{-1}, \quad (49)$$

where $N_{nuc} \sim 10^{48}$ is the number of nucleation centers expected in the same innermost region of the compact object discussed above (Iida & Sato 1998; Bombaci et al. 2016) and $\nu_0$ is the oscillation frequency computed in the WKB approximation. The details are reported in Iida & Sato (1997, 1998). In the nucleation time, the exponential dominates with respect to the prefactors. For that reason, in many works the prefactors are replaced by simple expressions as $T^4$ or $\mu^4$ (in units of s$^{-1}$ fm$^{-3}$; see, e.g., Di Toro et al. 2006; Mintz et al. 2010a, 2010b) due to dimensional arguments. In this paper, we will follow the latter approach and calculate nucleation times as

$$\tau^{\text{th}}(P_H, \{\Delta y_i\}, T) = [V_{\text{nuc}} \mu_{H^*}^4 \mathcal{P}_1 \text{th} \mathcal{P}_2 \text{th}]^{-1}[\text{s}], \quad (50)$$

$$\tau^{\text{qt}}(P_H, \{\Delta y_i\}, T) = [V_{\text{nuc}} \mu_{H^*}^4 \mathcal{P}_1^{\text{qt}} \mathcal{P}_2^{\text{qt}}]^{-1}[\text{s}]. \quad (51)$$

Of course, the real nucleation time is the minimum between the one calculated by using thermal nucleation or by using quantum nucleation:

$$\tau(P_H, \{\Delta y_i\}, T) = \min[\tau^{\text{qt}}, \tau^{\text{th}}]. \quad (52)$$

The idea is that sets of fluctuation $\{\Delta y_i\}$ can exist for which nucleation is more favorable than in the case in which only the average composition $\{y_i^{H_\beta}\}$ is considered. To allow such a process, however, it is necessary to "pay the price" of producing a fluctuation in the composition, $\{y_i^{H^*}\}$, whose probability strongly decreases as $\{\Delta y_i\}$ increases. Clearly, the larger the temperature, the more likely the fluctuations corresponding to large values of $\{\Delta y_i\}$. Thus, nucleation can be much more efficient with respect to the scenario considered in Bombaci et al. (2016). On the other hand, at small temperatures, $\mathcal{P}_1$ becomes vanishingly small except for very small values of $\{\Delta y_i\}$. The contribution of thermal fluctuations in the composition of the hadronic phase becomes negligible, and we thus expect to return to the case analyzed in Bombaci et al. (2016), in which the composition of the quark matter seed is identical, in terms of flavors, to the average composition of the hadronic phase at equilibrium $\{y_i^{H_\beta}\}$.

In principle, a complete discussion of the role of fluctuations should consider all the possible sets $\{\Delta y_i\}$, and the nucleation time should be computed by integrating over all those sets. However, this approach is computationally very time intensive. Therefore, in this work, we will focus only on two extreme cases. The first one, indicated with $\beta^*$, corresponds to $\{\Delta y_i = 0\}$ (i.e., no fluctuations in the hadronic composition, as done in Bombaci et al. 2016). The second one, $(\beta\beta)$, is based on a choice of $\{\Delta y_i\}$ such that $\{y_i^{Q^*}\} = \{y_i^{Q_\beta}\}$, namely in the hadronic subsystem, the flavor composition is identical to the flavor composition of quark matter in $\beta$-equilibrium. Thus, in the two-flavor case, the former $(\beta^*)$ corresponds to

$$y_p^{H^*}(P_H, T) = y_p^{H_\beta}(P_H, T), \quad (53)$$

$$y_n^{H^*}(P_H, T) = y_n^{H_\beta}(P_H, T), \quad (54)$$

$$y_e^{H^*}(P_H, T) = y_e^{H_\beta}(P_H, T), \quad (55)$$

while the latter $(\beta\beta)$ corresponds to

$$y_p^{H^*}(P_H, T) = \frac{2y_u^{Q_\beta}(P_H, T) - y_d^{Q_\beta}(P_H, T)}{3}, \quad (56)$$

$$y_n^{H^*}(P_H, T) = \frac{2y_d^{Q_\beta}(P_H, T) - y_u^{Q_\beta}(P_H, T)}{3}, \quad (57)$$

$$y_e^{H^*}(P_H, T) = y_e^{Q_\beta}(P_H, T), \quad (58)$$

where all the fractions are computed at fixed temperature and pressure. Let us motivate our choice of the $\beta\beta$ case as the most relevant over all possible fluctuations in $\{\Delta y_i\}$. The $(\beta\beta)$ case is the one corresponding to the minimal value of $\mu_{Q^*}(P_H, \{\Delta y_i\}, T)$, namely $\mu_{Q_\beta}$. By integrating over all possible sets of $\{\Delta y_i\}$, we expect that the probability associated to the $\beta\beta$ fluctuation is overwhelming (due to the exponential dependence of the probability of nucleation on $W_{2,c}$).

Finally, we specify that our approach, even in the ($\beta\beta$) case, is also different with respect to the one presented in Mintz et al. (2010a, 2010b). There, the authors include the effect of fluctuations in the hadronic composition by computing $W_2$ between the phases $H_\beta$ and $Q_\beta$, while $W_1 = 0$. On the other hand, in our approach, $W_2$ is computed between the phases $H^*$ and $Q^*$ (that is equivalent to $Q_\beta$ in the ($\beta\beta$) case), and $W_1$ is computed between the phases $H_\beta$ and $H^*$. In general, our approach in the ($\beta\beta$) case leads to a smaller critical radius (and thus a smaller $W_{2,c}$) than the one that can be obtained in Mintz et al. (2010a, 2010b), since $\mu_{H^*}(P, T) \geqslant \mu_{H_\beta}(P, T)$ (see Equations (32) and (33)), and thus a higher $\mathcal{P}_2$. However, in our approach, we still have the $\mathcal{P}_1$ contribution to be taken into account. At high temperature, we expect $\mathcal{P}_1$ to be high enough to make the nucleation faster within our approach. At low temperature, $\mathcal{P}_1$ decreases, and we thus expect a slower nucleation with respect to Mintz et al. (2010a, 2010b).

## 4. Equation of State

In this section, we report the EOSs used for nucleons, quarks, electrons, and respective antiparticles. All the models reported here are characterized by a single-particle energy spectrum (quasiparticle energy) for the species $j$ having the functional form[7]

$$\epsilon_{k_j} = (k_j^2 + m_j^2)^{1/2} + U_j(n_B, \{y_j\})$$
$$\equiv E_{k_j} + U_j(n_B, \{y_j\}). \quad (59)$$

It can be shown (C. Constantinou et al. 2024, in preparation) that the energy density of a particle or antiparticle species $j$ is

$$\varepsilon_j = \gamma_j \int_0^\infty \frac{d^3 k_j}{(2\pi)^3} \frac{E_{k_j}}{e^{(E_{k_j}-\mu_{K,j})/T}+1} + V_j, \quad (60)$$

where $\gamma_j$ is the degeneracy factor (spin, isospin, color, etc.). The first term is the Fermi integral of a free fermion gas that represents the kinetic term. The second term is the contribution of the interaction potential to the energy density. The kinetic chemical potential is $\mu_{K,j} = \mu_j - U_j$.

Similarly, the number densities are

$$n_j = y_j n_B = \gamma_j \int_0^\infty \frac{d^3 k_j}{(2\pi)^3} \frac{1}{e^{(E_{k_j}-\mu_{K,j})/T}+1}, \quad (61)$$

where $y_j$ is the particle fraction, i.e., the ratio between the number density $n_j$ of the particle or antiparticle species $j$ and the baryon density $n_B$. The pressure is

$$P_j = \gamma_j \int_0^\infty \frac{d^3 k_j}{(2\pi)^3} \frac{k_j^2}{E_{k_j}} \frac{1}{e^{(E_{k_j}-\mu_{K,j})/T}+1} +$$
$$+ n_B \left.\frac{\partial V_j}{\partial n_B}\right|_{\{y_j\}}, \quad (62)$$

where the first term is the kinetic contribution of particles or antiparticles, while the second term is the contribution of the interaction potential to the pressure. All the other thermodynamic quantities can be computed using the standard thermodynamic relations, e.g.,

$$\mu_j = \mu_{K,j} + \frac{1}{n_B}\left.\frac{\partial V_j}{\partial y_j}\right|_{n_B, \{y_{z\neq j}\}}, \quad (63)$$

$$s = \frac{1}{T}\left(\varepsilon + P + n_B \sum_j \mu_j y_j\right). \quad (64)$$

While at $T = 0$ the Fermi integrals can be computed analytically, the generalization to the finite-temperature case needs a numerical approach. In this work, we used the "JEL" numerical approach (Johns et al. 1996). Details of the numerical setup can be found in C. Constantinou et al. (2024, in preparation). Moreover, the temperature only affects the kinetic terms (i.e., the Fermi integrals) since we are considering models with a moment-independent potential-energy spectrum $U_j$. Particles and antiparticles will be considered in equilibrium, thus

$$\mu_j + \mu_{\bar{j}} = 0. \quad (65)$$

These conditions allow one to use as independent variables the net fractions (e.g., $y_e = y_{e^-} - y_{e^+}$). All the thermodynamic quantities have the form $X = X(n_B, \{y_i\}, T)$, where $X = P, \varepsilon, s, \mu_j$.

### 4.1. Nucleons

Hadrons will be described by the Zhao–Lattimer (ZL) EOS model (Zhao & Lattimer 2020; Constantinou et al. 2021). The ZL is a schematic nucleonic model based on the following energy density functional with an interaction potential between baryons that reads

$$V_H = V_p + V_n$$
$$= 4n_B^2 y_n y_p \left\{\frac{a_0}{n_0} + \frac{b_0}{n_0^\gamma}[n_B(y_n + y_p)]^{\gamma-1}\right\}$$
$$+ n_B^2 (y_n - y_p)^2 \left\{\frac{a_1}{n_0} + \frac{b_1}{n_0^{\gamma_1}}[n_B(y_n + y_p)]^{\gamma_1-1}\right\}, \quad (66)$$

where $n_0 \simeq 0.16 \text{ fm}^{-3}$ is the nuclear saturation density. The sum of the Fermi integrals of protons and neutrons gives the kinetic contributions. Although not based on microscopic physics, the interaction potential is similar to what can be obtained from a relativistic Lagrangian with vector interactions (but not scalar interactions) at the mean-field level. The ZL model allows one to reproduce several bulk nuclear properties by using only six parameters ($a_0$, $b_0$, $a_1$, $b_1$, $\gamma$, and $\gamma_1$). In particular, with a proper choice of parameters, this EOS is consistent with laboratory data at nuclear saturation density and with recent chiral effective field theory (CET) calculations up to $n_B < 2n_0$. Moreover, the ZL EOS can reproduce results that are consistent with recent astrophysical data (Constantinou et al. 2021).

### 4.2. Quarks

Quark dynamics will be described by the vMIT model (Klähn & Fischer 2015; Gomes et al. 2019; Constantinou et al. 2021). This differs from the traditional MIT bag model because the QCD perturbative terms are dropped and replaced by a repulsive vector interaction among quarks via the exchange of a vector-isoscalar meson. The potential contribution to the energy density is

$$V_Q = \sum_q V_q$$
$$= \frac{1}{2}a\left[n_B\left(\sum_q y_q\right)\right]^2 + B, \quad (67)$$

where $q = u, d, s, \bar{u}, \bar{d}, \bar{s}$ and $a = \left(\frac{G_V}{m_V}\right)^2$, where $G_V$ is the coupling constant and $m_V$ the mass of the meson. The sum of the $u, d, s, \bar{u}, \bar{d}, \bar{s}$ Fermi integrals gives the kinetic contributions for quarks.

---

[7] In the EOS section, we will call $j$ all the particles and antiparticles species separately and $i$ the net or total values as reported in footnote 6. That is, $j = p, n, u, d, s, \bar{u}, \bar{d}, \bar{s}, e^-, e^+$ and $i = p, n, u, d, s, e$. Antiprotons and antineutrons do not play a relevant role at $T \lesssim 100$ MeV, and thus they will not be considered.

### 4.3. Leptons

The leptons (electrons and positrons) are considered as a free Fermi gas. Thus, $V_{e^-} = V_{e^+} = 0$ and the thermodynamic quantities are expressed in terms of Fermi integrals.

### 4.4. Equilibrium Phases

The hadronic matter in $\beta$-equilibrium $H_\beta$ is characterized by

$$y_p^{H_\beta} + y_n^{H_\beta} = 1, \quad (68)$$

$$y_p^{H_\beta} - y_e^{H_\beta} = 0, \quad (69)$$

$$\mu_p^{H_\beta} + \mu_e^{H_\beta} = \mu_n^{H_\beta}, \quad (70)$$

namely the baryon number conservation, the charge neutrality, and the equilibrium with respect to $\beta$-reactions, where

$$\mu_h^{H_\beta} = \mu_h(n_B^{H_\beta}, \{y_h^{H_\beta}\}, T),$$
$$\mu_e^{H_\beta} = \mu_e(n_B^{H_\beta}, y_{e^-}^{H_\beta}, T),$$

where $h = p, n$.

The three-flavor quark matter in $\beta$-equilibrium, $Q_{\beta,3\text{flav}}$, is characterized by

$$\frac{1}{3}y_u^{Q_\beta} + \frac{1}{3}y_d^{Q_\beta} + \frac{1}{3}y_s^{Q_\beta} = 1, \quad (71)$$

$$\frac{2}{3}y_u^{Q_\beta} - \frac{1}{3}y_d^{Q_\beta} - \frac{1}{3}y_s^{Q_\beta} - y_e^{Q_\beta} = 0, \quad (72)$$

$$\mu_u^{Q_\beta} + \mu_e^{Q_\beta} = \mu_d^{Q_\beta}, \quad (73)$$

$$\mu_d^{Q_\beta} = \mu_s^{Q_\beta}, \quad (74)$$

namely the baryon number conservation, the charge neutrality, and the equilibrium with respect to all weak reactions, where

$$\mu_q^{Q_\beta} = \mu_q(n_B^{Q_\beta}, \{y_q^{Q_\beta}\}, T)$$
$$\mu_e^{Q_\beta} = \mu_e(n_B^{Q_\beta}, y_e^{Q_\beta}, T),$$

where $q = u, d, s$. In the two-flavor case, $Q_{\beta,2\text{flav}}$, Equation (74) is replaced by the condition $y_s^{Q_\beta} = 0$. Since in this work we are focused mainly on the two-flavor case, we will usually simply refer to it as $Q_\beta$.

The parameters used are reported in Table 1. In Appendix B, we show a comparison between the EOS models used in this work and some other possible models.

**Table 1**
Parameter Sets Used in the Present Work

| Model | Parameter | Value | Units |
|---|---|---|---|
| ZL | $a_0$ | −96.64 | MeV |
|  | $b_0$ | 58.85 | MeV |
|  | $\gamma$ | 1.40 | … |
|  | $a_1$ | −26.06 | MeV |
|  | $b_1$ | 7.34 | MeV |
|  | $\gamma_1$ | 2.45 | … |
| vMIT | $m_u$ | 5 | MeV |
|  | $m_d$ | 7 | MeV |
|  | $m_s$ | 150 | MeV |
|  | $a$ | 0.2 | fm$^2$ |
|  | $B^{1/4}$ | 165 | MeV |
| Constants | $\hbar c$ | 197.3 | MeV fm |
|  | $m_p, m_n$ | 939.5 | MeV |
|  | $m_e$ | 0.511 | MeV |

## 5. Results and Discussion

Figure 1 shows the Gibbs energy per baryon as a function of the pressure at $T = 20, 50$ MeV for hadronic matter in $\beta$-equilibrium $H_\beta$ (that is, $H^*(\beta^*)$), out-of-equilibrium hadronic matter $H^*(\beta\beta)$, two-flavor quark matter in $\beta$-equilibrium $Q_\beta$ (that is, $Q^*(\beta\beta)$), out-of-equilibrium quark matter $Q^*(\beta^*)$, and two-flavor quark matter in $\beta$-equilibrium $Q_{\beta,3\text{flav}}$. The definitions of these phases are reported in Section 3. At fixed pressure and temperature, the favored phase is the one with the lower Gibbs energy per baryon. As expected, the equilibrium phases ($H_\beta$ and $Q_\beta$) are favored with respect to the out-of-equilibrium phases ($H^*$ and $Q^*$, respectively). Moreover, the three-flavor quark matter in equilibrium is energetically favored with respect to the two-flavor due to the appearance of new degrees of freedom associated with strange quarks. At low pressure and temperature, the hadronic phase is stable. When the curve for the quark phase crosses the hadronic curve, the latter becomes metastable. In the $\beta\beta$ case, the hadronic phase becomes metastable with respect to the quark phase at $P \simeq 478$ MeV fm$^{-3}$ and at $P \simeq 427$ MeV fm$^{-3}$ for $T = 20$, 50 MeV, respectively. Instead, in the $\beta^*$ case, the hadronic phase becomes metastable at $P \simeq 1802$ MeV fm$^{-3}$ and $P \simeq 1709$ MeV fm$^{-3}$ for $T = 20, 50$ MeV, respectively. Thus, in the case a small hadronic subsystem is in a $H^*(\beta\beta)$ phase due to a thermal fluctuation, such a subsystem would become metastable with respect to the $Q_\beta$ quark phase at much lower pressures and temperatures in comparison to the ones at which a subsystem in the $H_\beta$ phase will become metastable with respect to $Q^*(\beta^*)$. Once the hadronic phase becomes metastable, fluctuations can generate a critical droplet of the new stable quark phase, leading then to deconfinement.

Moreover, Figure 1 also shows the flavors composition of $Q^*$ in the $\beta\beta$ and $\beta^*$ cases as a function of the pressure. In the two-flavor $Q_\beta$ phase (and thus in $Q^*(\beta\beta)$) the quark fractions are nearly constant, with $y_d^{Q_\beta} \sim 2y_u^{Q_\beta}$ and $y_e^{Q_\beta} \sim 0$. At the same time, the flavor composition in the $H_\beta$ phase (and thus in $Q^*(\beta^*)$) becomes more symmetric at high pressures due to the high symmetry energy at high densities obtained within the ZL EOS.

Figure 2 shows the mass–radius diagram and central pressure as a function of the mass of compact objects (purely hadronic stars, hybrid stars with Gibbs construction for two- and three-flavor quark phases). All the configurations with the used EOS models used support masses greater than $2\,M_\odot$. In the core of PNSs, with entropy per baryon $S = s/n_B = 2$, the pressure reaches $P \sim 51\,\text{MeV fm}^{-3}$ for $1.4\,M_\odot$ and $P \simeq 550\,\text{MeV fm}^{-3}$ for the maximum mass configuration $M_\text{max} \simeq 2.26\,M_\odot$. Note that the three-flavor $\beta$-equilibrium EOS will not play a role in nucleation calculations; it only describes (strange) quark matter at the end of the evolution, once the weak interaction has had sufficient time to minimize the energy of the system, changing the flavor composition via the reaction $u + d \to u + s$.

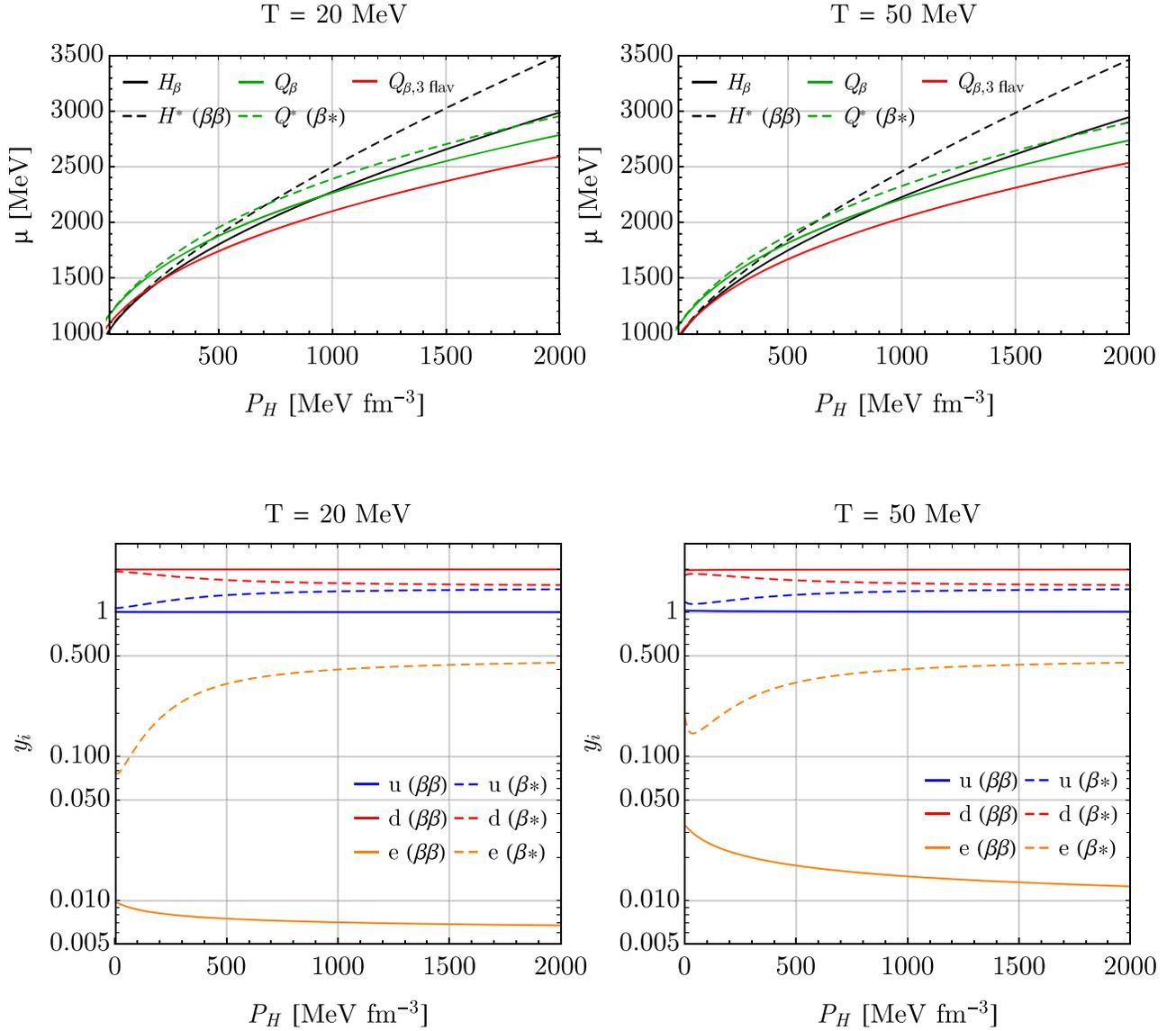

**Figure 1.** Upper panels: Gibbs energy per baryon vs. pressure for two values of the temperature ($T = 20$, $50$ MeV). The hadronic phase in $\beta$-equilibrium ($H_\beta$, equivalent to $H^*$ ($\beta^*$)), out-of-equilibrium hadronic phase ($H^*$ ($\beta\beta$)), two-flavor quark matter in $\beta$-equilibrium ($Q_\beta$, equivalent to $Q^*$ ($\beta\beta$)), out-of-equilibrium quark matter ($Q^*$ ($\beta^*$)), and three-flavor quark matter in $\beta$-equilibrium ($Q_{\beta,\mathrm{3flav}}$) are reported. Lower panels: flavor fractions of the quark phase vs. pressure for two values of the temperature ($T = 20$, $50$ MeV). The continuous and dashed lines refer to two-flavor quark matter for the case of $Q^*(\beta\beta)$, namely $Q_\beta$, and of $Q^*(\beta^*)$, respectively.

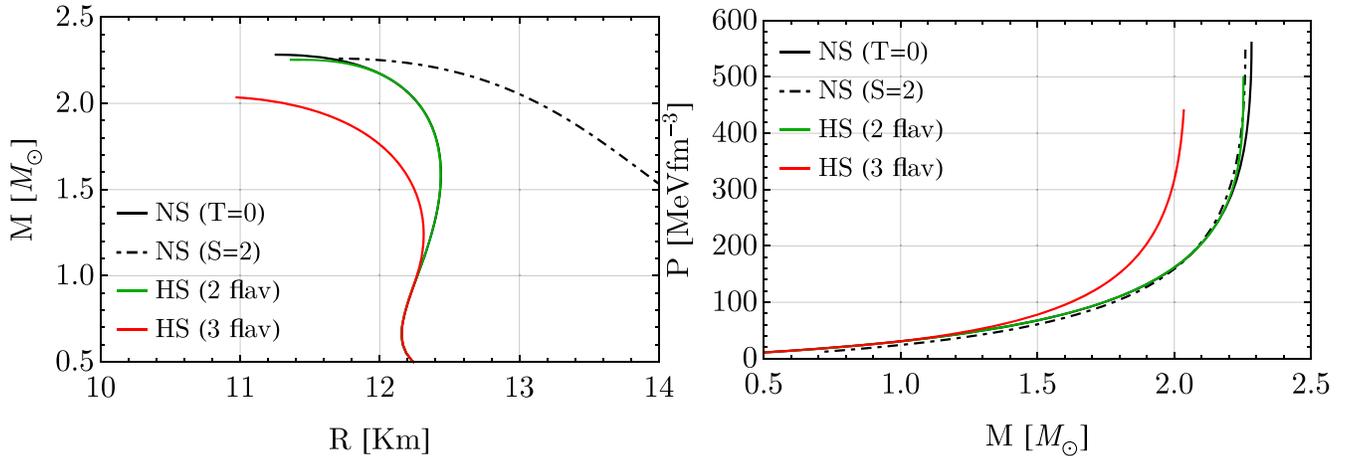

**Figure 2.** Mass–radius diagram (left) and central pressure vs. mass (right) for the pure hadronic EOS (NS) at $T = 0$ and $S = s/n_B = 2$, and for the hybrid hadron-quark EOS using a Gibbs construction using the two-flavor (HS 2 flav) and the three-flavor (HS 3 flav) quark phase at $T = 0$.

Figure 3 shows the probability of fluctuations (see Equation (40)) of a certain fraction of protons, neutrons, and (net) electrons $\Delta y_p$, $\Delta y_n = -\Delta y_p$, and $\Delta y_e = \Delta y_p$ with respect to the average $\beta$-equilibrium matter both in linear and logarithmic scale. The higher the temperature and pressure, and the lower the baryon number, the more probable the fluctuation of the composition with respect to the average values in a subsystem.

Figure 4 shows the works $W_1$ and $W_2$ for $\beta\beta$ and $\beta^*$ at $P = 470\,\text{MeV fm}^{-3}$ as a function of a droplet radius $R$ for $T = 20, 50\,\text{MeV}$ and $\sigma = 10, 30\,\text{MeV fm}^{-2}$. When the hadronic phase is stable, $W_2$ always increases as the radius increases. When the hadronic phase becomes metastable, $W_2$ reaches a maximum at the critical radius $R_c$, and can be interpreted as a finite potential barrier. The higher the pressure and temperature (and the lower the surface tension), the lower the potential barrier. In the $\beta^*$ case, the hadronic matter is stable in all the conditions reported in the plots. Thus, the corresponding work is constantly increasing as a function of the seed radius. In the $\beta\beta$ case, at $P = 470\,\text{MeV fm}^{-3}$, the hadronic phase is stable at $T = 20\,\text{MeV}$ and metastable at $T = 50\,\text{MeV}$. Thus, at $T = 20\,\text{MeV}$ the works $W_2$ are increasing as a function of the droplet radius, while at $T = 50\,\text{MeV}$ they reach a maximum. Finally, $W_1$ is always positive since it represents the energy cost needed to change the hadronic composition from the equilibrium.

Figure 5 shows the critical and turning point radius and the work at the critical radius as a function of the pressure. At the pressure and temperature at which the hadronic phase becomes metastable, the critical radius and the work $W_2$ at the critical radius diverge. By increasing the pressure and temperature, we obtain a smaller radius. The critical radius in the $\beta\beta$ case is much smaller than in $\beta^*$. At $P \gtrsim 577\,\text{MeV fm}^{-3}$ and $P \gtrsim 533\,\text{MeV fm}^{-3}$ for $T = 20, 50\,\text{MeV}$, respectively, the turning points radii are equal to the critical radius $R_- = R_+ = R_c$. That means that the WKB semiclassical ground state's energy is equal to the maximum of the potential barrier. Thus, at fixed temperature, at pressures higher than the ones reported above, the action under the barrier (Equation (36)) is zero, and the probability of quantum tunneling through the barrier $\mathcal{P}_2^\text{qt}$ becomes 1. Thus, at such conditions, the only contribution to the total quantum nucleation probability is given by $\mathcal{P}_1^\text{qt}$. Finally, a much lower critical radius in $\beta\beta$ leads to a much lower critical work $W_2(R_c)$ (i.e., the maximum of the potential barrier). At the reported temperatures, the work $W_1(R_c)$ ($\beta\beta$) is greater than $W_2(R_c)$ ($\beta\beta$) but much lower than $W_2(R_c)$ ($\beta^*$). This is a first indication that by allowing the hadronic composition to fluctuate, the gain in nucleation probability from a convenient out-of-equilibrium subsystem of the hadronic phase is greater than "the cost to be paid" to have such a subsystem with composition different form the average ones.

Figure 6 shows the number of baryons in a critical radius and turning point radius droplet as functions of the pressure and for different values of the temperature. These baryon numbers identify the hadronic subsystems in which the composition differs from the average values. The figure also displays the difference in terms of the number of protons in the subsystem to obtain the case $\beta\beta$. Obviously, $\Delta N_p = 0$ in $\beta^*$. Let us focus, for example, on the set of parameters $\sigma = 30\,\text{MeV fm}^{-2}$, $P = 500\,\text{MeV fm}^{-2}$, and $T = 50\,\text{MeV}$. In a critical radius seed, there are $\sim 43$ baryons; to obtain the $\beta\beta$ configuration in this subsystem, $\sim 13$ protons should be replaced by neutrons, namely the proton fraction should be reduced by $\sim 0.3$. The probability of such a fluctuation in the hadronic composition is very low, $\mathcal{P}_1 \sim 10^{-35}$, but, as we will see later, not low enough to compensate for the huge advantage of nucleating the quark phase starting from a $\beta\beta$ subsystem with respect to a $\beta^*$ one.

Figure 7 shows the nucleation time as a function of the pressure for different temperatures in the $\beta\beta$ and $\beta^*$ cases. It is useful to note that the nucleation time varies very quickly with pressure and temperature. For instance, at $T = 50\,\text{MeV}$, a pressure difference of $\sim 20\,\text{MeV fm}^{-3}$ is enough to vary the nucleation time by a hundred orders of magnitude (thus much larger than the age of the Universe, $\sim 10^{17}\,\text{s}$).

Within the temperature conditions shown in the plots, the nucleation time in the $\beta\beta$ case is much shorter than in the $\beta^*$ case at fixed temperature and pressure (or, equivalently, for the same temperature and nucleation time, the pressure in the $\beta\beta$ case is much lower). Thus, the role of fluctuations in hadronic phase composition is crucial. In the $\beta^*$ case, thermal nucleation dominates over quantum nucleation at high temperatures. By decreasing the temperature, the thermal nucleation time increases faster than the quantum nucleation time, which in turn becomes almost temperature independent (note that the blue, green, and black dashed lines in the left plot are very close). Thus, the quantum nucleation is dominant in the low-temperature regime. In particular, at a temperature of $T \simeq 5.4\,\text{MeV}$, the curves corresponding to the quantum and to the thermal nucleation times cross at $\tau \sim 1\,\text{s}$. We can then consider this "crossover temperature" as the one separating the thermal and the quantum nucleation regimes (at $\sigma = 30\,\text{MeV fm}^{-2}$). In the $\beta\beta$ case, the crossover temperature is $\simeq 7.8\,\text{MeV}$. Again, thermal and quantum nucleation dominate at high and low temperatures, respectively.

Notice that quantum nucleation remains temperature dependent even at low temperatures and large pressures, since under such conditions $R_c = R_+$ and, thus, the quantum tunneling probability $\mathcal{P}_2^\text{qt}$ is 1, but the probability of a thermal fluctuation (of the hadronic phase composition) $\mathcal{P}_1^\text{qt}$ remains temperature dependent. This feature corresponds to the change in the slope of the quantum nucleation time for pressures larger than $P_H \sim 578\,\text{MeV fm}^{-3}$ (see the green dashed line in Figure 7).

In Figure 8, the thermodynamic conditions (pressure and temperature) at which the nucleation time is 1 s are shown for the $\beta\beta$ and $\beta^*$ cases. The two-flavor Gibbs mixed-phase boundaries are also reported for comparison. Notice that the choice of 1 s for the nucleation time is just arbitrary. Indeed, as shown before, for a fixed value of the surface tension, one could obtain a nucleation time varying in a very wide range, $\sim (10^{-20}$–$10^{17})\,\text{s}$, by slightly changing the pressure and temperature conditions as reported above. Our main result is that, according to what was shown before, the $\beta\beta$ case allows the nucleation of quark matter at much lower pressures and temperatures than the $\beta^*$ case (namely the case in which fluctuations are not considered). Also, thermal nucleation at high temperatures is faster than the quantum tunneling nucleation, as expected. By decreasing the temperature, the thermal nucleation is suppressed faster than the quantum nucleation, since in the former the temperature appears explicitly in the exponential, while in the latter the temperature is only (implicitly) present in $W_2$. As explained before, while at temperatures below $\sim 10$ MeV the quantum nucleation of the $\beta^*$ case is almost temperature independent, the quantum nucleation of the $\beta\beta$ case has two components: $\mathcal{P}_2^{qt}$ is nearly temperature independent (and, in particular, is $\mathcal{P}_2^{qt} \sim 1$ if $R_+ = R_c$, namely for $P_H \gtrsim 512, 580$ MeV fm$^{-3}$ for $\sigma = 10, 30$ MeV fm$^{-2}$, respectively), but $\mathcal{P}_1$, (i.e., the probability of finding a subsystem with the composition $\beta\beta$) drops as the temperature decreases. The role of the fluctuations in the hadronic composition becomes negligible for $T \lesssim 0.1$, 1 keV for $\sigma = 10$, 30 MeV fm$^{-2}$, respectively. As expected, the higher the surface tension, the higher the pressure needed to start the nucleation process at a fixed temperature, the larger the difference between thermal and quantum nucleation, and the larger the temperature at which hadronic composition fluctuations become negligible. The fluctuations in the hadronic composition thus substantially increase the nucleation efficiency in the high- and intermediate-temperature regimes.

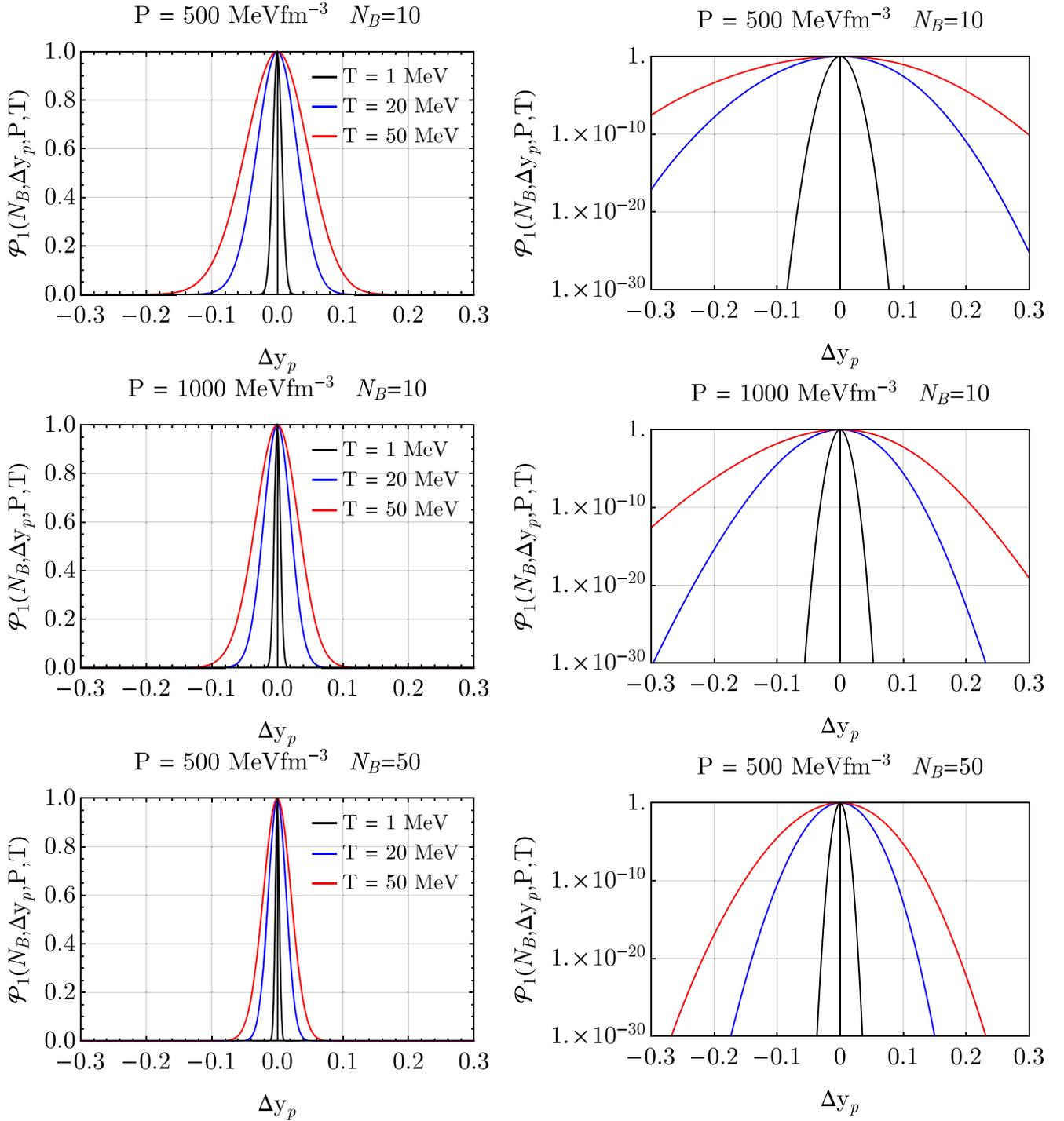

**Figure 3.** Probability (not normalized) that at fixed pressure $P$, temperature $T$, and baryon number $N_B$, the net fraction of protons, neutrons, and electrons differs from the average ones by $\Delta y_p$ (notice that $\Delta y_n = -\Delta y_p$ and $\Delta y_e = \Delta y_p$). Both linear (left) and logarithmic (right) scales are shown. The three curves correspond to different values of $T$ at $P = 500$ MeV fm$^{-3}$ and $N_B = 10$ (first row), $P = 1000$ MeV fm$^{-3}$ and $N_B = 10$ (second row), and $P = 500$ MeV fm$^{-3}$ and $N_B = 50$ (third row).

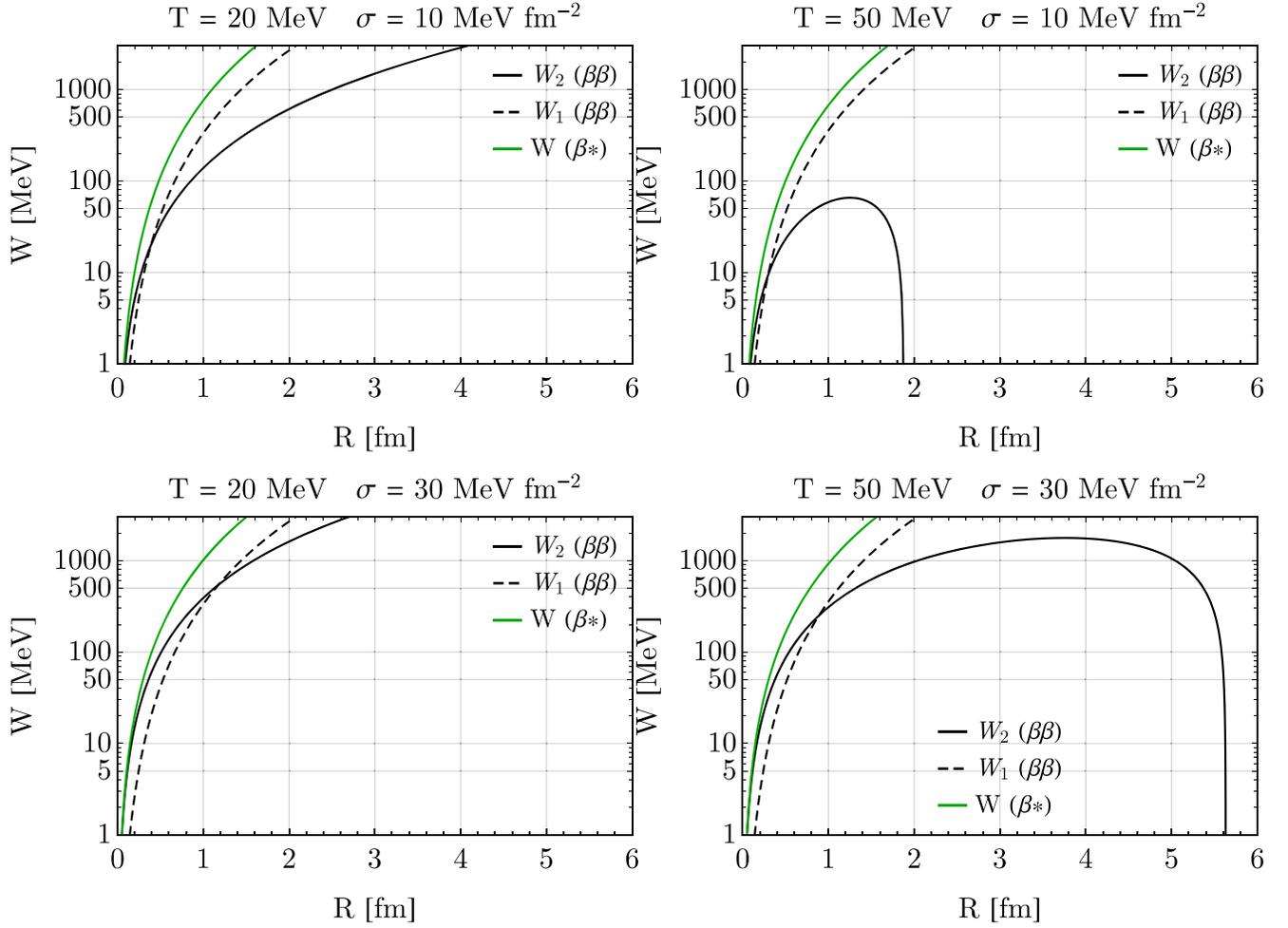

**Figure 4.** Work vs. quark droplet radius at $T = 20$ MeV (left) and $T = 50$ MeV (right) and $\sigma = 10$ MeV fm$^{-2}$ (top) and $\sigma = 30$ MeV fm$^{-2}$ (bottom) in the $\beta\beta$ and $\beta^*$ cases. For $\beta\beta$ both $W_1$ and $W_2$ are reported, while for $\beta^*$ $W = W_2$, $W_1 = 0$. The pressure is fixed at $P = 470$ MeV fm$^{-3}$.

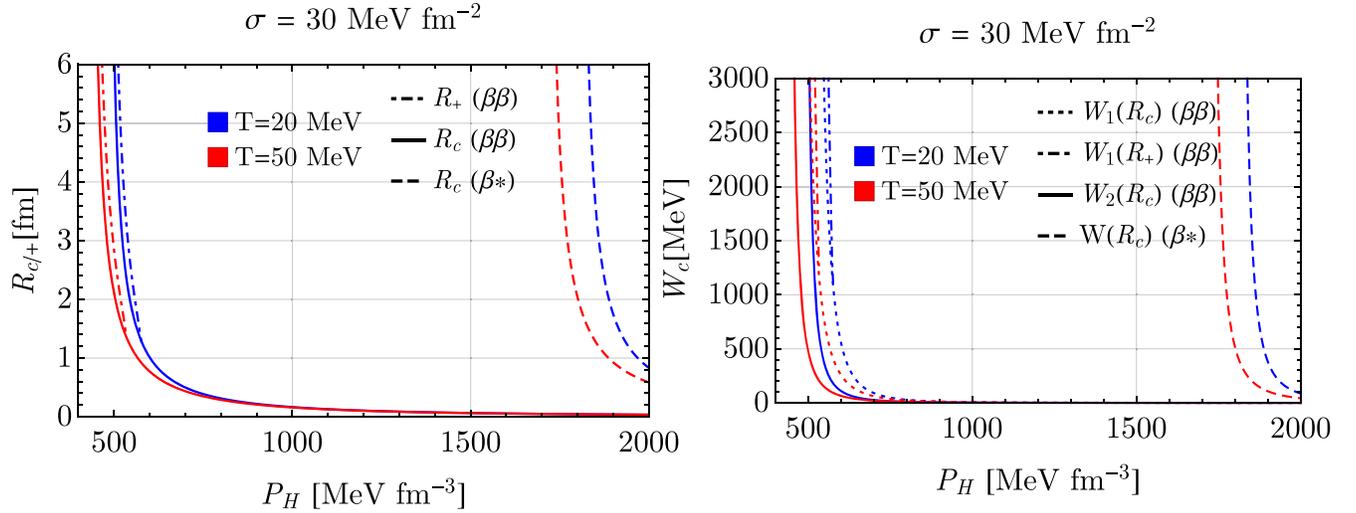

**Figure 5.** Critical $R_c$ and turning point $R_+$ radius (left) and work at the critical radius (right) for the $\beta\beta$ and $\beta^*$ cases.

Let us discuss now the normalization factor $K(P_H, T)$ introduced in Section 3.2. At fixed pressure, temperature,

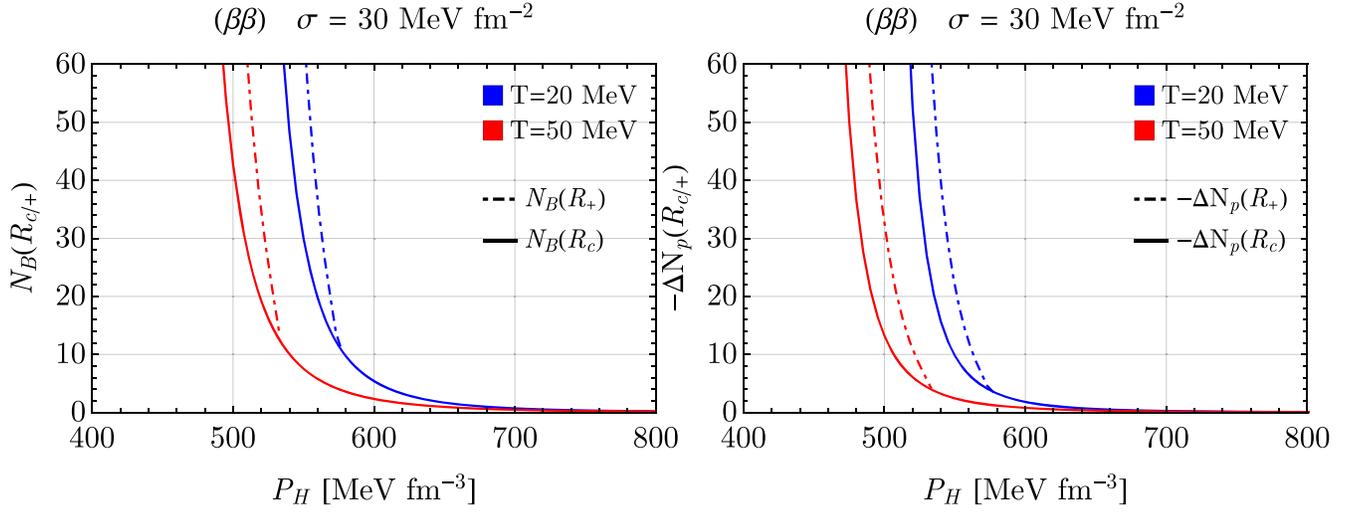

**Figure 6.** Baryon number (left) and fluctuation in the number of protons (right) in a droplet of quark matter with a critical $R_c$ or turning point $R_+$ radius in the $\beta\beta$ case.

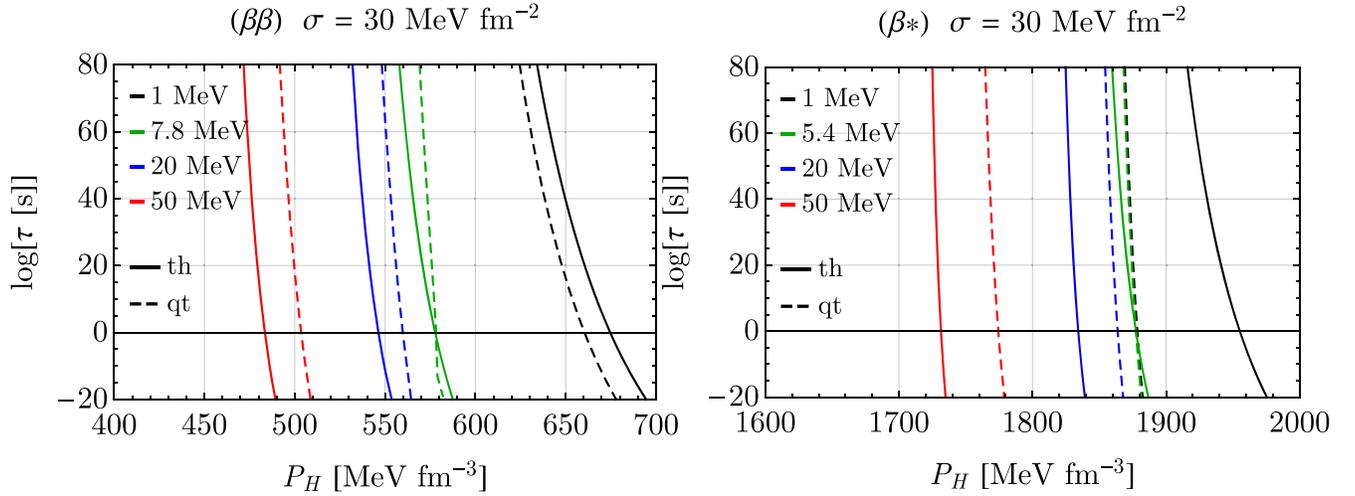

**Figure 7.** Logarithm of (thermal and quantum) nucleation time vs. pressure in the $\beta\beta$ (left) and $\beta*$ (right) cases. In both cases the temperatures $T = 1, 20, 50$ MeV are shown. Moreover, $T \simeq 7.8, 5.4$ MeV are reported to show the conditions at which quantum nucleation becomes faster in the plotted range for $\beta\beta$ and $\beta^*$, respectively.

number of baryons, and surface tension, $K(P_H, T)$ is the inverse of the integral of the curves shown in Figure 3. By numerically calculating these integrals along the pressure and temperature curves corresponding to $\tau = 1$ s shown in Figure 8, we note that $K(P_H, T)$ turns out to be approximately constant and of the order of ~20. It can be seen from Figure 7 that a constant shift of an order of magnitude in nucleation time, all other conditions being equal, would not have a qualitative impact. In particular, Figure 8 would not show qualitatively significant changes. Thus, we consider the approximation $K(P_H, T) = 1$ to be reliable.

Let us now consider the typical thermodynamic conditions (of pressure and temperature) that are realized in compact stars. Consider, for example, a newly born hadronic PNS formed following a CCSN. Qualitatively, we can approximate the thermal evolution of the PNS as follows (Prakash et al. 1997). About ~0.5 s after the explosion, the PNS is characterized by an entropy per baryon of about ~1−3, and neutrinos are in thermodynamic equilibrium with the system (neutrino trapping), resulting in a lepton fraction of $Y_{Le} \simeq 0.4$. At this stage, nucleation is suppressed due to the aforementioned neutrino trapping, as found in Bombaci et al. (2016). After ~(5−10) s, the PNS is deleptonized: Neutrinos are no longer in thermodynamic equilibrium with the rest of the system (i.e., the neutrino mean free path is much larger than the size of the PNS) and they freely stream out of the star. Thus, at this stage, the matter of the PNS is approximately in $\beta$-equilibrium, the chemical potential of the neutrinos is zero, and the entropy per baryon reaches a value of about ~2. During the next ~60 s, the PNS rapidly cools down by emitting neutrinos and the core temperature drops to a few MeV (see, e.g., Pons et al. 1999). By comparing the nucleation curves in Figure 8 with the purple curve (i.e., the $s/n_B = 2$ neutrino-less hadronic PNS profile) of the same figure, one can notice that a PNS with a central pressure $P \gtrsim 465, 505$ MeV fm$^{-3}$ (with corresponding mass ~2.26 $M_\odot$) can nucleate quark matter (assuming $\sigma = 10, 30$ MeV fm$^{-2}$, respectively).

Another interesting scenario in which our scheme can be applied corresponds to the situation in which deconfined quark matter is produced in a failed CCSN explosion (see, e.g., Fischer et al. 2018). In their analysis, they assume that a stable mixed phase is generated without any delay. The approach by

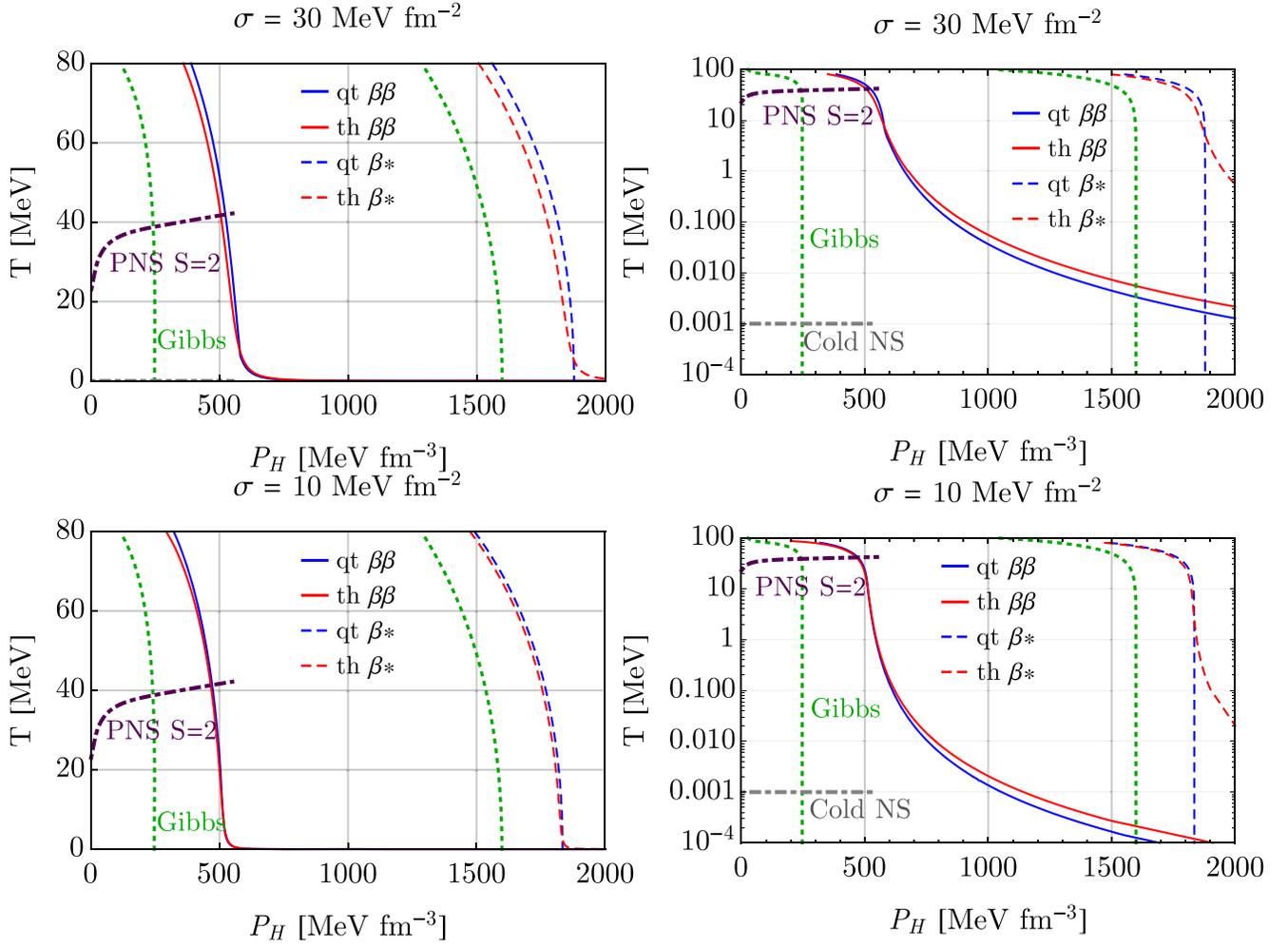

**Figure 8.** Temperature and pressure (of the hadronic phase) at which the (thermal and quantum) nucleation time is 1 s. Both the linear (left) and logarithmic (right) scales for $\sigma = 30$ MeV fm$^{-2}$ (top) and $\sigma = 10$ MeV fm$^{-2}$ (bottom) are shown. The thermal and quantum nucleation are reported for the $\beta\beta$ and $\beta^*$ cases. The mixed-phase boundaries of the two-flavor Gibbs construction are reported for comparison. The purple dotted–dashed curve in the right panel represents the pressure and temperature of the core of PNSs, assuming a $s/n_B = 2$ hadronic and neutrino-free matter (i.e., approximately the conditions 10–60 s after the core collapse). The end point of this curve corresponds to the PNS maximum mass configuration. The gray dotted–dashed curve in the right panels indicates the range of pressures reached in the core of cold NSs (assumed to have a uniform temperature $T = 1$ keV).

Bombaci et al. (2016, corresponding to the $\beta^*$ case), does not allow the formation of quark matter before reaching densities so large that the collapse to a black hole is triggered. Instead, taking into account the composition fluctuations, the formation of the first droplet of quark matter is delayed, but the needed overpressure is not so large as to be associated with the collapse to a black hole.

In contrast, nucleation cannot occur in a cold NS when only two flavors are considered.

## 6. Conclusions

The main contribution of this work has been to set a new framework for the study of the nucleation of quark matter within metastable hadronic matter at finite temperature which takes account the thermal fluctuations in the hadronic composition. Indeed, a standard recipe is to consider $\beta$-stable hadronic matter and to impose the condition that the flavor content of the newly formed quark matter droplets is equal to the one of the initial hadronic phase, due to the different timescales associated with the weak and strong interactions. However, at finite temperatures, the unavoidable occurrence of thermal fluctuations of the hadronic composition can lead to a faster nucleation process. We have taken into account these effects by computing the nucleation probability as the product between two terms:

1. The probability that in a hadronic system at chemical equilibrium, a small subsystem containing a certain baryon number is in a phase $H^*$ characterized by a composition that differs from the average one by a set of fluctuations $\{\Delta y_i\}$.
2. The probability to nucleate from the subsystem $H^*$ a quark seed $Q^*$ having the same flavor composition.

In this first work, we have limited the discussion to the case of two-flavor quark matter (thus only nucleons in the hadronic phase and up and down in the quark phase). A complete discussion of the presented framework should consider all the possible sets of fluctuations $\{\Delta y_i\}$ and the nucleation time should be computed by integrating over all these sets. However, in this work, we focused on two extreme cases, namely $\beta^*$, characterized by $\{y_i^{Q^*}\} = \{y_i^{H^*}\} = \{y_i^{H_\beta}\}$ (i.e., no fluctuations in the hadronic composition $\{\Delta y_i\} = 0$) and $\beta\beta$, characterized by $\{y_i^{Q^*}\} = \{y_i^{H^*}\} = \{y_i^{Q_\beta}\}$ (i.e., the hadronic flavor composition fluctuations are chosen in order to match the $\beta$-equilibrium quark composition).

Our results could be important for numerical studies on the formation of quark matter in astrophysical processes. Indeed, it is common to assume that quark matter is produced in equilibrium (mechanical, chemical, and thermal) with the hadronic phase; that is, quark matter is promptly formed once the central density of the star reaches the critical density for the formation of the mixed phase, within the Gibbs construction (Sagert et al. 2009; Fischer et al. 2018). This approach neglects finite-size effects (namely the nucleation process) that would unavoidably delay the formation of quark matter. While in the scheme presented in Bombaci et al. (2016, corresponding to the $\beta^*$ case), the delay could be sizable enough to prevent the formation of quark matter in astrophysical systems (at least in the two-flavor case), in the scheme here developed the effective threshold for the appearance of quark matter is still reachable in compact stars under specific conditions. In particular:

1. At low temperatures $T \lesssim (0.1-1)$ keV, the role of fluctuations of the hadronic composition is totally negligible and our results do not differ from those of Bombaci et al. (2016).
2. At intermediate temperatures $(0.1-1)$ keV $\lesssim T \lesssim (1-10)$ MeV, the fluctuations of the hadronic composition increase the efficiency of nucleation but, at least for the EOSs discussed in this work, the pressure needed to nucleate cannot be reached in (cold) compact stars.
3. At high temperatures $T \gtrsim (1-10)$ MeV, the process of nucleation can take place in the pressure regime of astrophysical phenomena related to compact objects (e.g., very massive PNSs, CCSNe, or BNSMs).

We emphasize again that our work has focused only on the two-flavor case (specifically nucleons in the hadronic phase and $u$ and $d$ quarks in the quark phase, using specific EOS models for the different phases) and thus does not consider any kind of interaction concerning strangeness. In a forthcoming paper, we will also include hyperons (and thus strange quarks) and investigate the effect of the nucleation process on the phenomenology of the two-families scenario, namely the scenario in which hadronic stars and strange quark stars coexist (Drago et al. 2014, 2016; Drago & Pagliara 2016). Similarly, color superconductivity and boson condensates are not included in this work and their effects will be investigated in the future. It is important to remark that the formalism here developed for dealing with nucleation in multicomponent systems could also be applied in other physical systems in which a phase transition to or from quark matter occurs, as, for instance, in heavy ions collisions (Di Toro et al. 2006; Bonanno et al. 2007).

## Appendix A
## Fluctuations as a Multivariate Gaussian

It is interesting to note that in the small fluctuation limit ($\{\Delta y_i\} \ll 1$, $\{\Delta N_i\} \ll N_B$), $\exp(-W_1/T)$ is a multivariate Gaussian. Let us expand, up to second order,

$\Phi_{H^*}(P_H, \{N_i^{H^*}\}, T)$ for $\{N_i^{H^*}\} \simeq \{N_i^{H_\beta}\}$:

$$\Phi_{H^*}(P_H, \{N_i^{H^*}\}, T) \simeq \Phi_H(P_H, \{N_i^{H_\beta}\}, T)+$$
$$+ \sum_i \left.\frac{\partial \Phi_H}{\partial N_i}\right|_{P_H, T, N_i = N_i^{H_\beta}} \Delta N_i +$$
$$+ \sum_{i,j} \left.\frac{\partial^2 \Phi_H}{\partial N_i \partial N_j}\right|_{P_H, T, N_i = N_i^{H_\beta}} \Delta N_i \Delta N_j$$
$$= \Phi_H(P_H, \{N_i^{H_\beta}\}, T)+$$
$$+ \sum_i \mu_i^{H_\beta} \Delta N_i + \frac{1}{2}\sum_{i,j} \left.\frac{\partial \mu_i^{H_\beta}}{\partial N_j^{H_\beta}}\right|_{P_H, N_{z \neq j}, T} \Delta N_i \Delta N_j.$$

By replacing in Equation (40), we obtain

$$W_1 \simeq \frac{1}{2} \sum_{i,j} \left.\frac{\partial \mu_i^{H_\beta}}{\partial N_j^{H_\beta}}\right|_{P_H, N_{z \neq j}, T} \Delta N_i \Delta N_j. \quad (A1)$$

Note that Equation (A1) does not contain any term linear in $\Delta N_i$. This is expected since we are considering fluctuations around the state of minimal energy. Thus,

$$\exp\left[-\frac{W_1}{T}\right] \simeq \prod_{i,j} \exp[-\beta_{ij} \Delta N_i \Delta N_j], \quad (A2)$$

where

$$\beta_{ij} = \frac{1}{2T} \left.\frac{\partial \mu_i^{H_\beta}}{\partial N_j^{H_\beta}}\right|_{P_H, N_{z \neq j}, T}. \quad (A3)$$

One can note that Equation (A2) indeed corresponds to a multivariate Gaussian distribution. In order to estimate the contribution to the probability given by this dimensionless normalization factor, let us consider the multivariate Gaussian reported in Equation (A2). The normalization factor of a multivariate Gaussian is analytically known (Landau & Lifshitz 1969):

$$K(P_H, T) \simeq \frac{\sqrt{\det|\beta|}}{(2\pi)^{m/2}}, \quad (A4)$$

where $m$ is the number of flavors (or conserved charges) considered and $\det|\beta|$ is the determinant of the matrix having as elements $\beta_{ij}$, as defined in Equation (A3).

## Appendix B
## More on EOS Models

In Figure 9, we report some more information regarding the used EOS models (see Section 4). In the left panel, we display the pressure as a function of the energy density for the ZL model, for the mixed phase of the Gibbs construction (see Constantinou et al. 2021) between the ZL and vMIT models, for the vMIT and for the simple bag model (i.e., vMIT with $a = 0$) in the case of cold matter in $\beta$-equilibrium. The gray shaded area represents the region excluded by Komoltsev & Kurkela (2022) using theoretical considerations arising from CET and perturbative QCD (pQCD). The used models are compatible with such constraints in the plotted region. The models are no longer compatible with such constraints at much higher pressure and energy density. Notice, however, that the vMIT model provides an EOS for quark matter suitable for the typical thermodynamical conditions reached in compact objects (i.e., approximately up to the black dots shown in the plot, which represent the conditions in the center of the maximum mass configurations of NSs and hybrid stars). At the same time, it does not aim to describe deconfined quark matter up to the perturbative regime (while, for example, models such as Fraga et al.'s 2014 are built to be compatible with the perturbative regime). The density-dependent repulsive interaction between quarks in the vMIT model makes the EOS much stiffer with respect to the one obtained within the simple bag model, allowing $\gtrsim 2 \, M_\odot$ hybrid stars. In the right panel, the pressure as a function of the baryon density for pure nucleonic symmetric matter in the ZL model is compared with the SFHO EOS model (Steiner et al. 2013), and it is shown to be compatible with the constraints from heavy-ion collisions provided by Danielewicz et al. (2002).[8] Moreover, the ZL model with the parameterization used is compatible with the CET calculations available at $n_B \lesssim 2n_0$ and reported in Drischler et al. (2021), and with the nuclei properties at zero temperature as shown in Constantinou et al. (2021).

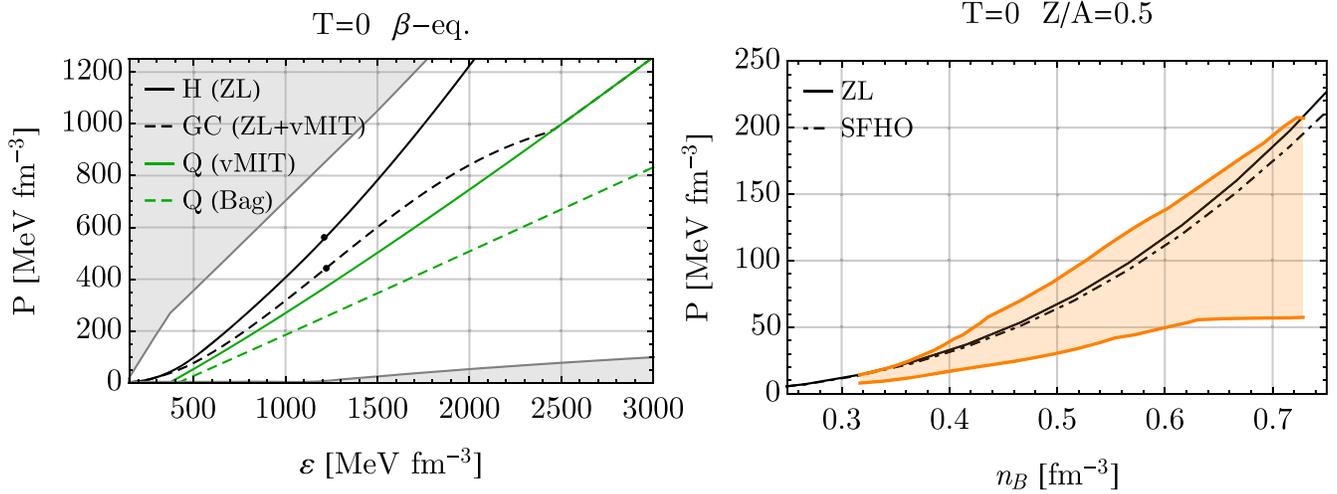

**Figure 9.** Left panel: pressure vs. energy density for neutrino-less cold matter in $\beta$-equilibrium (i.e., NS matter). The pure nucleonic ZL case, the Gibbs construction (GC) mixed phase between nucleonic matter and three-flavor quark matter, the pure three-flavor quark phase using the vMIT model, and a simple bag model are reported. The black points identify the pressure and the energy density in the center of the maximum mass configurations in the cold NS and hybrid star. The gray shaded area represents the theoretical constraints to the energy density and pressure arising from CET and pQCD (Komoltsev & Kurkela 2022). Right panel: pressure vs. baryon density for cold symmetric matter. The used hadronic EOS ZL is compared with the SFHO EOS (Steiner et al. 2013). The orange shaded area shows the baryon density and pressure region compatible with the nuclear collision experiments (Danielewicz et al. 2002). The mixed quark-hadron phase does not appear in the plot since for two-flavor matter it starts at very large baryon densities ($n_B \simeq 1.58$ fm$^{-3}$).

## ORCID iDs

Mirco Guerrini ⓘ https://orcid.org/0009-0009-5946-3900
Giuseppe Pagliara ⓘ https://orcid.org/0000-0003-3250-1398
Alessandro Drago ⓘ https://orcid.org/0000-0003-1302-8566
Andrea Lavagno ⓘ https://orcid.org/0000-0001-7675-304X

---

[8] https://compose.obspm.fr/eos/33